\documentclass[final]{siamltex}
\usepackage{graphicx}
\usepackage{morefloats} 
\usepackage{color}
\usepackage{amsmath}
\usepackage{amsfonts}
\usepackage{amssymb}
\usepackage{hyperref}
\usepackage[colorinlistoftodos]{todonotes}
\usepackage{algorithm}
\usepackage{algorithmic}
\usepackage{fix-cm}
\usepackage[normalem]{ulem}
\usepackage{url}
\usepackage{verbatim}
\usepackage{float}
\usepackage{epstopdf}

\newtheorem{remark}{Remark}

\newtheorem{example}{Example}

\newcommand{\vect}[1]{\boldsymbol{#1}}  
\newcommand\fc{\vect{f}_{\!c}}
\newcommand\M{\vect{\mathcal{M}}}

\title{Constrained Runs algorithm as a lifting operator for the Boltzmann equation}

\author{Ynte Vanderhoydonc\thanks{Dept.~Mathematics and Computer Science, Universiteit Antwerpen, Middelheimlaan 1, 2020 Antwerpen, Belgium ({\tt ynte.vanderhoydonc@uantwerp.be}).}
        \and
        Wim Vanroose\thanks{Dept.~Mathematics and Computer Science, Universiteit Antwerpen, Middelheimlaan 1, 2020 Antwerpen, Belgium ({\tt wim.vanroose@uantwerp.be}).}}

\begin{document}

\maketitle

%
%

\begin{abstract}
Lifting operators play an important role in starting a kinetic Boltzmann model from given macroscopic information. The macroscopic variables need to be mapped
to the distribution functions, mesoscopic variables of the Boltzmann model.
A well-known numerical method for the initialization of Boltzmann models is the Constrained Runs algorithm. This algorithm is used in literature for the initialization of lattice Boltzmann models, special discretizations of the Boltzmann equation. It is based on the attraction of the dynamics toward the slow manifold and uses lattice Boltzmann steps to converge to the desired dynamics on the slow manifold. We focus on applying the Constrained Runs algorithm to map density, average flow velocity, and temperature, the macroscopic variables, to distribution functions. Furthermore, we do not consider only lattice Boltzmann models. We want to perform the algorithm for different discretizations of the Boltzmann equation and consider a standard finite volume discretization.
\end{abstract}




\section{Introduction}

Fluid dynamics can typically be described by different levels of
accuracy. One distinguishes between micro-, meso-, and macroscopic
scales. Macroscopic partial differential equations (PDEs) model only a
few low order velocity moments and are therefore not that accurate to
describe interactions between particles. Well-known macroscopic
equations for the modeling of fluid dynamics are the Euler and
Navier--Stokes equations. However, when a detailed description is
necessary, a microscopic model is used. These individual-based models
take the particle collision physics into account. The Boltzmann
equation can be used to describe kinetic models, that are ubiquitous
to model at this scale, by modeling distribution functions in phase
space.

We focus on space, time, and velocity discretizations of the Boltzmann
equation. The initialization of such a discretized Boltzmann equation
requires a lifting operator when only macroscopic information is
available. This operator defines a mapping between macroscopic and
microscopic/mesoscopic variables such that the distribution functions
of the discretized Boltzmann equation can be built from the given
macroscopic information. The concept of a lifting operator in a
multiscale context was introduced by Kevrekidis et al.~in the
equation-free framework to couple different scales in a dynamical
system \cite{kevrekidis}. Similar ideas were introduced in the
heterogeneous multiscale methods framework that starts from a
predetermined incomplete form for the macroscopic model and estimates
the needed data for the incomplete macroscopic model from the
microscopic model \cite{heterogeneous}.

Lots of previous work in the literature is based on the initialization
of lattice Boltzmann models (LBMs), special discretizations of the
Boltzmann equation \cite{succi, LGCA_LBM}.  Lifting operators for LBMs
are considered in \cite{leemput_phd, leemput, initialization,
  vandekerckhove, vanderhoydonc1, vanderhoydonc2, eBook} where a given
initial density is mapped to distribution functions. The
Chapman--Enskog expansion is built for such model problems in
\cite{leemput_phd}. This analytical expansion is first introduced in
\cite{chapman_cowling} to solve the Boltzmann equation. Furthermore,
\cite{leemput_phd}, \cite{leemput}, \cite{initialization}, and
\cite{vandekerckhove} apply the Constrained Runs (CR) algorithm to
LBMs. Originally, the CR algorithm is introduced by Gear et
al.~\cite{gear} to map macroscopic initial variables to missing
microscopic variables for stiff singularly perturbed ordinary
differential equations (ODEs). This algorithm is based on the
attraction of the dynamics toward the slow manifold. The dynamics on
this slow manifold can be parameterized by only macroscopic variables
such as the density. The higher order velocity moments become slaved
functionals of the density in the LBM context. \cite{vanderhoydonc1} compares the Chapman--Enskog expansion
with the CR algorithm for hybrid model problems with given initial
density. \cite{vanderhoydonc2} constructs the numerical
Chapman--Enskog expansion as a lifting operator for these model
problems. This numerical Chapman--Enskog expansion is based on a
combination of the Chapman--Enskog expansion and the CR algorithm. It
finds the coefficients of the Chapman--Enskog expansion numerically
based on the CR algorithm. This reduces the number of unknowns in the
lifting since it only finds the coefficients of the expansion rather
than the full state of distribution functions. A review of these
lifting operators for LBMs with an initial density as macroscopic
variable is given in \cite{eBook}. A generalization of the numerical
Chapman--Enskog expansion is made in \cite{vanderhoydonc3} where both
density and momentum are considered to be given at initialization.

In this paper, we apply the Constrained Runs algorithm on a finite
volume discretization of the one-dimensional Boltzmann equation which
maps macroscopic information, namely density, average flow velocity,
and temperature, to distribution functions of the discretized
Boltzmann equation.  The intention of this generalization is to deal
with multiple given macroscopic variables. We do not conserve only the
density but also average flow velocity and temperature to work towards
more realistic physical systems. This lifting operator is constructed
to deal, for example, with missing data in a hybrid kinetic and
macroscopic PDE model for laser ablation. In laser ablation, a complex
interplay between various processes determines the outcome of a laser
beam hitting on a solid target \cite{bogaerts}. First, the target is
heated by the laser and the heat is transported over the material. The
material then starts to melt.  Above the surface of the melted
material there is evaporation and the particles escape according to a
certain velocity distribution. Due to the strong evaporation, the
particles right above the melt/gas interface are not in an equilibrium
state.  This thin layer above the surface is known as the Knudsen
layer which cannot be described by a macroscopic PDE. It requires a
detailed kinetic model such as the Boltzmann equation. Once away from
the surface the plume expansion can be described by a macroscopic PDE
model.  The challenge is to couple the models describing the different
physical processes in a mathematical correct way.

We first applied the generalization of the CR algorithm in
\cite{vanderhoydonc3} for a small number of velocities to deal with
the conservation of density and momentum in lattice Boltzmann
problems.

The outline of this paper is as follows. Section \ref{description}
contains the different levels of description that will be used
throughout the paper. The mesoscopic level uses the discretized
version of the Boltzmann equation while the macroscopic equivalent
PDEs are the Euler and Navier--Stokes equations. Section
\ref{constrained_runs} describes the Constrained Runs algorithm. The
origin of the Constrained Runs algorithm is included in section
\ref{origin_CR}. Section \ref{CR_LBMs} contains the application of
Constrained Runs to lattice Boltzmann models. The generalization to
different discretizations of the Boltzmann equation and multiple
conserved moments is outlined in section
\ref{generalization_CR}. Numerical results can be found in section
\ref{numerical_results} where the lifting operator is tested in a
setting of restriction and lifting. A conclusion and outlook are
included in section \ref{conclusion}.


\section{Different levels of description} \label{description} This
section introduces models with different levels of description that
will be used throughout this paper. The mesoscopic scale is described
by a finite volume discretization of the Boltzmann equation and will
be discussed in section \ref{mesoscopic}. The macroscopic scale uses
discretized partial differential equations (PDEs) to describe the evolution of a
few low order velocity moments in section \ref{macro_eq}.


\subsection{Mesoscopic description} \label{mesoscopic} 
Kinetic models
make use of the Boltzmann equation \cite{succi} that describes the
evolution of a distribution function $f(\vect{x},\vect{v},t)$
(function space $C^2_{\mathbb{R}}(D)$) that counts the number of
particles or individuals in point $\vect{x} \in D_{\vect{x}}\subset
\mathbb{R}^n$, $n \in \mathbb{N}_0$, with a velocity $\vect{v} \in
D_{\vect{v}} \subset \mathbb{R}^n$, at time $t \geq 0$.  The equation
is given by (without external forces, with Bhatnagar--Gross--Krook
(BGK) collision term \cite{BGK})
\begin{equation}\label{boltzmann}
\frac{\partial}{\partial t}f(\vect{x},\vect{v},t)+\vect{v} \cdot \frac{\partial}{\partial \vect{x}} f(\vect{x},\vect{v},t) = \omega(f^{eq}(\vect{x},\vect{v},t)-f(\vect{x},\vect{v},t)).
\end{equation}
 The Maxwell--Boltzmann equilibrium distribution is
\begin{equation*}
f^{eq}(\vect{x},\vect{v},t)=n\left(\frac{m}{2 \pi k_B T}\right)^{d/2}e^{\left(-\dfrac{m(\vect{v}-\vect{u})^2}{2 k_B T}\right)},
\end{equation*}
where $k_B$ is the Boltzmann constant, $d$ the number of spatial dimensions, $m$ the molecular mass, $n$ the number density, $T$ the temperature, and $\vect{u}$ the average flow velocity. The macroscopic variables are defined by \cite{struchtrup}
\begin{eqnarray} \label{lower_moments}
n(\vect{x},t)&=&\int f(\vect{x},\vect{v},t)\mathrm{d}\vect{v}, \nonumber \\
\rho(\vect{x},t)&=&\int m f(\vect{x},\vect{v},t)\mathrm{d}\vect{v}, \nonumber \\
\rho(\vect{x},t) \vect{u}(\vect{x},t)&=&\int m \vect{v}f(\vect{x},\vect{v},t) \mathrm{d}\vect{v}, \nonumber \\
T(\vect{x},t) &=& \frac{1}{d\rho R} \int m \|\vect{v}-\vect{u}\|^2 f(\vect{x},\vect{v},t) \mathrm{d}\vect{v},
\end{eqnarray}
with $\|.\|$ the two-norm and $R=k_B/m$ the specific gas constant.

The BGK collision term approximation represents a relaxation towards equilibrium with a relaxation parameter $\omega$ and an associated time scale $\tau = 1/\omega$. Atomic collisions are taken into account by the relaxation frequency $\omega$ \cite{aoki} given by
\begin{equation}\label{omega_gusarov}
\omega=\rho\frac{k_B}{m}\frac{T}{\mu},
\end{equation}
with $\mu$ the viscosity of the gas which is determined by 
\begin{equation*}
\mu = \mu_\text{ref} \left(\frac{T}{T_\text{ref}}\right)^{\omega_\mu}.
\end{equation*}
$\omega_\mu$ represents the viscosity index of the considered gas \cite{bird}, and $\mu_\text{ref}$ and $T_\text{ref}$ are the reference viscosity and temperature determined experimentally. 

Consider the one-dimensional Boltzmann equation in space
\begin{equation*}
\frac{\partial f}{\partial t}+ v \frac{\partial f}{\partial x}=\omega(f^{eq}-f).
\end{equation*}
The velocity discretization of the Boltzmann equation results into a set of $N_v$ linear advection equations, with $N_v$ the number of velocity directions and $v_i=v_0+i\Delta v$, $i\in \{0,\ldots, N_v-1\}$ and $\Delta v = (v_\text{max}-v_\text{min})/N_v$.
\begin{equation} \label{adv_equations_boltzmann}
\frac{\partial f_{i}}{\partial t} + v_i \frac{\partial f_{i}}{\partial x} = \omega(f_i^{eq}-f_i).
\end{equation}

The discretization of the equilibrium distribution function is outlined below and is based on the algorithm described in \cite{gusarov}.
We look for the equilibrium distribution function $f_i^{eq}$ in the general form
\begin{equation}\label{eq_1D}
f^{eq}_{i}=A \exp \left(-B^2(v_i-D)^2 \right),
\end{equation}
which has to approach the Maxwell--Boltzmann distribution as the increment $\Delta v$ tends to zero.
$A$, $B$, and $D$ are derived from the conservation equations of mass, momentum, and energy, which are defined by
\begin{align*}
\Delta v \sum_{i=0}^{N_v-1}f_i&=n=\Delta v \sum_{i=0}^{N_v-1}f^{eq}_i(A,B,D), \\
\Delta v \sum_{i=0}^{N_v-1}v_i f_i &=n u = \Delta v \sum_{i=0}^{N_v-1}v_i f^{eq}_i(A,B,D), \\
\Delta v \sum_{i=0}^{N_v-1}(v_i-u)^2 f_i&=\frac{n k_BT}{m}  =  \Delta v \sum_{i=0}^{N_v-1}(v_i-u)^2 f^{eq}_i(A,B,D). \\
\end{align*}
From this follows
\begin{equation*}
A = \frac{n}{R_0}, \quad 
R_1 = 0, \quad
R_2 - \frac{ R_0 k_B T}{m} = 0,
\end{equation*}
where 
\begin{equation*}
R_j = \Delta v \sum_{i=0}^{N_v-1}(v_i-u)^j \exp \left( -B^2(v_i-D)^2 \right),
\end{equation*}
from which $A$, $B$, and $D$ are obtained through a Newton algorithm on this set of equations
with the initial approximation of $B=\sqrt{m/(2k_B T)}$ and $D=u$.

For the space and time discretization, we consider the one discussed in \cite{lafitte} by Lafitte and Samaey. 
It represents a finite volume discretization of the Boltzmann equation.
Consider a grid which is uniform in time with time step $\Delta t$ and a spatial grid with space step $\Delta x$:
\begin{equation*}
C_j=[x_{j-\frac{1}{2}},x_{j+\frac{1}{2}}), \quad 1 \leq j \leq N,
\end{equation*}
centered in $x_j$ with $x_j=j\Delta x$,
\begin{equation*}
T_k=[t^k,t^{k+1}), \quad k \geq 0, \quad t^k=k\Delta t.
\end{equation*}
By integrating \eqref{adv_equations_boltzmann} on a cell $M_{j,k}=C_j \times T_k$, this results in 
\begin{equation*}
\begin{split}
f_i(x_j,t^{k+1})= &f_i(x_j,t^k)-\frac{\Delta t}{\Delta x} \left(\phi(f)_{i,j+1/2}^k-\phi(f)_{i,j-1/2}^k\right) \\
 &+\Delta t \omega \left(f_i^{eq}(x_j,t^k)-f_i(x_j,t^k)\right),
 \end{split}
\end{equation*}
with upwind fluxes
\begin{equation*}
\phi_u(f)_{i,j+1/2}^k=
\begin{cases}
v_i f_i(x_j,t^k), \quad v_i \geq 0, \\
v_i f_i(x_{j+1},t^k), \quad v_i < 0,
\end{cases}
\end{equation*}
or centered fluxes
\begin{equation*}
\phi_c(f)_{i,j+1/2}^k=v_i \frac{f_i(x_{j+1},t^k)+f_i(x_{j},t^k)}{2}.
\end{equation*}

Different discretizations of the Boltzmann equation are given in \cite{mieussens}. It represents a similar explicit finite volume scheme and a linearized implicit scheme to compute steady states. It focuses on the positivity of solutions, the conservation of moments, and the dissipation of entropy.


\subsection{Macroscopic description} \label{macro_eq}

When the phenomena of fluid dynamics that one wants to describe are
macroscopic, the fluid is regarded as a continuous medium. The
fundamental equations to describe the motion of the fluid are the
Euler and Navier--Stokes equations \cite{chorin}. It can be shown that
these are macroscopic equivalents of the Boltzmann equation. For this,
the Chapman--Enskog expansion can be used \cite{chapman_cowling,
  LGCA_LBM}.

The Euler equations represent conservation of mass, momentum, and
energy while the Navier--Stokes equations extends these equations to
include the viscosity of the fluid.


\subsection{Initialization}
Suppose there is some initial data based on macroscopic variables,
like density, average flow velocity, and temperature. An important
question is how to start simulating a Boltzmann model, which is based
on distribution functions, given initial macroscopic
information. Starting the Boltzmann scheme includes some arbitrariness
since there are many possible distribution functions which have the
same velocity moments. We need to create a mapping from the
macroscopic velocity moments to distribution functions. For this a
lifting operator is necessary. Section \ref{constrained_runs}
describes the Constrained Runs algorithm that is used in this paper as
a lifting operator for the initialization of Boltzmann models. It is a
known algorithm for the initialization of lattice Boltzmann
models. This paper generalizes it to apply it for different
discretizations of the Boltzmann equation and multiple conserved
velocity moments.


\section{Constrained Runs algorithm} \label{constrained_runs} The
Constrained Runs (CR) algorithm is a numerical method based on the
attraction of the dynamics toward the slow manifold. The origin of the
CR algorithm in systems of ODEs is described in section
\ref{origin_CR}. The application to LBMs with given initial density is
included in section \ref{CR_LBMs}. Here, we immediately present the
generalized CR algorithm for LBM problems. The generalization to
different discretizations of the Boltzmann equation and multiple
conserved moments (density, average flow velocity, and temperature) is
discussed in section \ref{generalization_CR}.


\subsection{Origin of Constrained Runs algorithm} \label{origin_CR}
The CR algorithm finds its origin in systems of ODEs \cite{gear}. A
review of this work is given in \cite{eBook} and summarized
below. Given system
\begin{eqnarray} \label{system_ODEs}
\frac{\mathrm{d} \vect{r}(t)}{\mathrm{d} t}=p(\vect{r}(t),\vect{s}(t)), \nonumber \\
\frac{\mathrm{d} \vect{s}(t)}{\mathrm{d} t}=q(\vect{r}(t),\vect{s}(t)),
\end{eqnarray}
where only the initial condition for $\vect{r}$, namely
$\vect{r}(0)=\vect{r}_{\!0}$, is given. The aim is to find
$\vect{s}(0)=\vect{s}_0$ such that the initial condition
$(\vect{r}_{\!0},\vect{s}_0)$ lies on (or close to) the slow
manifold. The latter can be formulated by the function
$\vect{s}_0=\vect{s}(\vect{r}_{\!0})$.

Gear et al.~\cite{gear} proposed to obtain the $\vect{s}$-value from
equation
\begin{equation}\label{voorwaarde_slow} 
\frac{\mathrm{d}^{m+1}\vect{s}(t=0)}{\mathrm{d}t^{m+1}}=0,  
\end{equation}
the smoothness condition, with $m=0,1,\ldots$ that can be approximated by a forward difference
\begin{equation}\label{forward_form}
\Delta^{m+1}\vect{s}(t) \approx \Delta t^{m+1} \frac{\mathrm{d}^{m+1}\vect{s}(t)}{\mathrm{d}t^{m+1}}.
\end{equation}
It can be shown that this difference approximation used in the CR
algorithm can be interpreted as a backward extrapolation
\cite{vandekerckhove_phd}. It corresponds with a backward
extrapolation in time based on a polynomial of degree $m$ that passes
through the values $\vect{s}_k$ with $k = 1, \ldots ,m+1$ while the
known variable $\vect{r}$ is reset to its original initial value
$\vect{r}_{\!0}$.  The used coefficients of the forward finite
difference formulas at time $t$ are listed in Table
\ref{coeff_forward} for different degrees of $m$.

\begin{table}[!htop] 
\begin{center} 
\begin{tabular}{|c | c| c| c| c|}\hline
$m$ & $t+\Delta t$ & $t+2\Delta t$ & $t+3\Delta t$ & $t+4\Delta t$ \\ \hline
0 & 1 & 0 & 0 & 0\\
1 & 2 & $-1$ & 0 & 0\\
2 & 3 & $-3$ & 1 & 0\\
3 & 4 & $-6$ & 4 & $-1$\\ \hline
\end{tabular}
\end{center}
\caption{Coefficients of the forward finite difference formulas at time $t$ for different degrees of $m$ in \eqref{forward_form}. For example, for $m=1$, this formula corresponds to $\Delta t^2 \frac{\mathrm{d}^2 \vect{s}(t)}{\mathrm{d}t^2} \approx \vect{s}(t)-2\vect{s}(t+\Delta t)+\vect{s}(t+2\Delta t)$. Using \eqref{voorwaarde_slow}, this leads to $\vect{s}(t)=2\vect{s}(t+\Delta t)-\vect{s}(t+2\Delta t)$ from which the coefficients can be obtained. \label{coeff_forward}}
\end{table}

The general CR algorithm for a constant extrapolation, $m=0$, is given in Algorithm \ref{alg:CR_ODE}.
\begin{algorithm}[!htop]
\caption{Constrained Runs for a constant extrapolation in time in the system of ODEs \eqref{system_ODEs} \label{alg:CR_ODE}}
\begin{algorithmic}
\REQUIRE Initial condition $\vect{r}(0)=\vect{r}_{\!0}$ \\
Choose $\vect{s}_0$, norm $\|.\|$ and a tolerance $\theta$ \\
\REPEAT
\STATE Advance the model with  one time step $\Delta t$: \\
\quad $\vect{r}_{\!1}$ and $\vect{s}_1$ at time $t=\Delta t$ \\
\STATE Difference approximation $\Delta \vect{s}_0=\vect{s}_1-\vect{s}_0$ \\
\STATE $\vect{s}_0 \leftarrow \vect{s}_0+\Delta \vect{s}_0$ \\
\STATE Reset $\vect{r}$ to $\vect{r}_{\!0}$, the given initial condition \\
\UNTIL $\|\Delta \vect{s}_0 \| < \theta$
\end{algorithmic}
\end{algorithm}

Similar algorithms can be constructed for higher degrees of $m$ by
advancing the model during more time steps and using
\eqref{voorwaarde_slow} for different values of $m$.


\subsection{Constrained Runs algorithm applied to lattice Boltzmann models} \label{CR_LBMs}

The CR scheme \cite{gear} is a fixed point iteration scheme that
computes the full state of a microscopic time simulator on (or close
to) the slow manifold corresponding to the given macroscopic
variables.

Consider an LBM problem in a D1Q3 setting (one spatial dimension,
three velocity directions $v_i=c_i \Delta x / \Delta t$, $c_i=i$,
$i\in \{-1,0,1\}$).  The lattice Boltzmann equation (LBE) is
\begin{equation*}
f_i(x+c_i\Delta x,t+\Delta t)=(1-\omega)f_i(x,t)+\omega f_i^{eq}(x,t).
\end{equation*}
This is a special discretization of the Boltzmann equation
\eqref{boltzmann} \cite{succi, LGCA_LBM}.  The equilibrium
distributions are given by $f_i^{eq}(x,t)=\frac{1}{3}\rho(x,t)$, $i
\in \{-1,0,1\}$ \cite{vandersman} in which the particle density
$\rho(x,t)$ is defined as the zeroth order moment of the distribution
functions $\rho(x,t)=\sum_{i \in \{-1,0,1\}} f_i(x,t)$. These
equilibrium distributions correspond to a local diffusive equilibrium.

We now describe the application of Constrained Runs to LBMs
\cite{leemput_phd,initialization} when we assume that a slow manifold
exists in this context.

The CR algorithm sets a few LBM steps after which the density is
reset. The number of LBM steps is related to the order $m$ and
determines the accuracy of the resulting lifted distribution
function. Doing only one LBM step corresponds to a constant
extrapolation in time.

The CR procedure for LBMs iterates upon the higher order moments
$\vect{\phi}$ and $\vect{\xi}$, momentum and energy, given a density
$\vect{\rho}_0 = \rho(\vect{x},0)$. It is equivalent to
determine$\vect{f} :=
\{\vect{f}_{\!1};\vect{f}_{\!0};\vect{f}_{\!-1}\}$ or $\M :=
\{\vect{\rho};\vect{\phi};\vect{\xi}\}$ since the velocity moments are
related  via moment matrix $M \in \mathbb{R}^{3 \times 3}$:
\begin{equation}\label{omzet}
\left(\begin{array}{c}
\rho \\
\phi \\
\xi
\end{array}\right)
=
\left(\begin{array}{c c c}
1 & 1 & 1 \\
1 & 0 & -1 \\
\frac{1}{2} & 0 & \frac{1}{2}
\end{array}\right)
\left(\begin{array}{c}
f_{1} \\
f_0 \\
f_{-1}
\end{array}\right)
=M
\left(\begin{array}{c}
f_{1} \\
f_0 \\
f_{-1}
\end{array}\right),
\end{equation}
or, in shorthand,  $\M = M \vect{f}$. 
Denote $M^0$ as 
\begin{equation*}
M^0 = \left(\begin{array}{c c c}
1 & 1 & 1 \\
0 & 0 & 0 \\
0 & 0 & 0
\end{array}\right) = \text{diag}(1,0,0) M.
\end{equation*}
$M^0$ represents the part of matrix $M$ that produces the density which should be conserved in the CR algorithm.

To apply CR, the missing moments are written as
\begin{equation*}
\vect{s}=\left(\begin{array}{c}\vect{\phi} \\ \vect{\xi} \end{array}\right),
\end{equation*}
a long vector $\vect{s} \in \mathbb{R}^{2N}$, the variable
$\vect{r}_{\!0}=\vect{\rho}_0 \in \mathbb{R}^{N}$ denotes the known
initial condition, with $N$ the number of spatial grid points.

The vector $\vect{s}^{(k)}$ denotes the $k$-th iterate of the CR
algorithm and the iterations are related by
\begin{equation} 
\vect{s}^{(k+1)}=\mathcal{C}_m\left(\vect{r}_{\!0},\vect{s}^{(k)}\right), \label{eq_CR}
\end{equation}
where $\mathcal{C}_m$ denotes one step of the CR algorithm and $m$ is
related to the order of the time derivative that is set to zero in
equation \eqref{voorwaarde_slow}.

Below, this paper presents the CR algorithm in a general setting.
We write the CR iteration in terms of the distribution
functions $\vect{f}$ rather than missing moments $\vect{\phi}$ and
$\vect{\rho}$.  The algorithm generates a sequence  of
moments $\M^{(k)}$ with $k=0,1,2,\ldots$ and $\M^{(0)} = \{\vect{\rho}_0; \left(\frac{1}{3}-\frac{1}{3}\right)\vect{\rho}_0;\frac{1}{2}\left(\frac{1}{3}+\frac{1}{3}\right)\vect{\rho}_0\}$
the initial guess corresponding to the equilibrium distribution function. In this sequence, the zeroth moment $\M^{(k)}_0$ is
always the initial density $\vect{\rho}_0$ and $\M^{(k)}_1 =
\vect{\phi}^{(k)}$ and $\M^{(k)}_2 =\vect{\xi}^{(k)}$ converge to the
slaved state.  The iteration consists of two steps. First, $\M^{(k)}$
is transformed to $\vect{f}^{(k)}$ using \eqref{omzet} and this is
used as an initial state of an LBM that is evolved over $m+1$ time steps.
This gives states $\vect{f}^{(k)}(\Delta t)$, $\vect{f}^{(k)}(2 \Delta
t)$, \ldots, $\vect{f}^{(k)}((m+1) \Delta t)$ that can be converted to
$\M^{(k)}(\Delta t)$, $\M^{(k)}(2\Delta t)$, \ldots,
$\M^{(k)}((m+1)\Delta t)$.  With these states we can generate
\begin{equation}
  \M_i^{(k)\text{pre}} = \sum_{j=1}^{m+1} w_j \M_i^{(k)}(j \Delta t)  \quad  \text{for} \quad i=1,2,
\end{equation}
the improved higher order moments where we add a superscript ``pre''
to denote that this is the value before resetting the density.  The underlying
idea is that $\M^{(k)\text{pre}}$ is chosen such that combined with
$\M^{(k)}(j\Delta t)$ it satisfies the smoothness condition
\eqref{voorwaarde_slow} by letting the weights $w_j$ correspond to the coefficients of the
forward finite difference formula in \eqref{forward_form} given in Table \ref{coeff_forward}.

For the ease of analysis, we also apply the backward interpolation formula
to the zeroth moment. Now, we can write for all moments that
\begin{equation}
  \M^{(k)\text{pre}} = \sum_{j=1}^{m+1} w_j \M^{(k)}(j \Delta t).  
\end{equation}
To get the next iterate we introduce a reset step that sets the zeroth moment to $\vect{\rho}_0$ such that the density is conserved. 
\begin{equation*}
  \M^{(k+1)}  = \begin{pmatrix} \vect{\rho}_0 \\ \M^{(k)\text{pre}}_1 \\ \M^{(k)\text{pre}}_2
  \end{pmatrix}.
\end{equation*}
This reset is the second step in the iteration.  The two steps can be written as
\begin{align*}
  \M^{(k+1)} &= \begin{pmatrix}
      0 & 0 & 0  \\
      0 & 1 & 0  \\
      0 & 0 & 1 
    \end{pmatrix}
    \M^{(k)\text{pre}} +  
\begin{pmatrix}
1  &0&0\\
0&0&0\\
0&0&0\\
\end{pmatrix}
\begin{pmatrix}
\vect{\rho}_0\\0\\0
\end{pmatrix},
\\
&= \left(M-M^0\right) \vect{f}^{(k)\text{pre}} +    
M^0  \vect{f}^0,
\end{align*}
where $\vect{f}^0$ is an initial distribution function that has $\vect{\rho}_0$ as density. Reordering leads to
\begin{align*}
  M \vect{f}^{(k+1)} &=   \left(M-M^0\right)\vect{f}^{(k)\text{pre}} + M^0 \vect{f}^0,   \\
  \vect{f}^{(k+1)} &= M^{-1 }\left[ \left(M-M^0\right)\vect{f}^{(k)\text{pre}} + M^0 \vect{f}^0 \right], \\
  \vect{f}^{(k+1)} &= \left(I-M^{-1}M^0\right)\vect{f}^{(k)\text{pre}}   + M^{-1}M^0 \vect{f}^0.
\end{align*}
In conclusion we can write 
\begin{equation}\label{eq:CR_LBM}
\vect{f}^{(k+1)} = \left(I-M^{-1}M^0\right) \left(\sum_{j=1}^{m+1}w_j \vect{f}^{(k)}(j\Delta t)\right) + M^{-1}M^0 \vect{f}^0.
\end{equation}
The convergence rate and stability of this iteration
depend on the order $m$ and the coefficients $w_j$ from Table
\ref{coeff_forward}.  

Note that the operator $P:=I-M^{-1}M^0$ that appears in the first term of \eqref{eq:CR_LBM}
is a projection operator.  Indeed, we can write $M^{-1}M^0 =
M^{-1}$diag$(1,0,0) M$ and it is now easy to see that $(M^{-1}M^0)^2 =
M^{-1} $diag$(1,0,0) M M^{-1} $diag$(1,0,0) M = M^{-1} M^0$.  As a
result $P^2=P$.

The interpretation of \eqref{eq:CR_LBM} is that the first term
projects the state on the space orthogonal to the space spanned by the
lowest order moments.  The second term adds the original
components in the space of the low order moments, here the density.

In general, equation \eqref{eq_CR} is nonlinear and
the fixed point can be found by a Newton iteration \cite{vandekerckhove}.  This means
solving
\begin{equation} \label{g_CR}
g_m(\vect{r}_{\!0},\vect{s}):=\vect{s}-\mathcal{C}_m(\vect{r}_{\!0},\vect{s})=0,
\end{equation}
for a given macroscopic value $\vect{r}_{\!0}$.
Newton's method gives an update to the guesses as follows
\begin{equation*}
\vect{s}^{(k+1)}=\vect{s}^{(k)} + \delta \vect{s}^{(k)},
\end{equation*}
where the corrections $\delta \vect{s}^{(k)}$ are found by solving the linear system
\begin{align*} 
A\left(\vect{r}_{\!0},\vect{s}^{(k)}\right) \delta \vect{s}^{(k)} &= \frac{\partial g_m}{\partial \vect{s}} \left(\vect{r}_{\!0},\vect{s}^{(k)}\right)  \delta \vect{s}^{(k)}, \nonumber \\
&= \left(I-\frac{\partial \mathcal{C}_m}{\partial \vect{s}}\left(\vect{r}_{\!0},\vect{s}^{(k)}\right)\right)  \delta \vect{s}^{(k)},  \nonumber \\
&= -g_m\left(\vect{r}_{\!0},\vect{s}^{(k)}\right),
\end{align*}
with $A=\frac{\partial g_m}{\partial \vect{s}}$ the linearization (Jacobian
matrix) of $g_m$ and $\frac{\partial \mathcal{C}_m}{\partial \vect{s}}$ the
linearization of $\mathcal{C}_m$. $I$ represents the identity matrix. The linearization can be estimated with the help of the approximation
\begin{equation} \label{num_jac}
A\vect{e}_i \approx \frac{g_m(\vect{r}_{\!0},\vect{s}+\varepsilon \vect{e}_i)-g_m(\vect{r}_{\!0},\vect{s})}{\varepsilon},
\end{equation}
with $\vect{e}_i$ the unit vector, $i=1,\ldots, 2n$ and $\varepsilon$ small.


\subsection{Constrained Runs algorithm applied to general discrete Boltzmann equations} \label{generalization_CR}
The previous section outlined the application of the Constrained Runs
algorithm to a lattice Boltzmann model with three velocities and
where only the density, the lowest order velocity moment, is given.

One question is how the method should be applied to other
discretizations of the Boltzmann equation, for example the finite
volume discretization discussed in section \ref{mesoscopic}.  These
discrete models have many more discrete velocities than the LBM. Consider
again a one-dimensional spatial domain. Instead of three, we have
$N_v$ velocity directions with velocities represented as
$\{v_0,\ldots,v_{N_v-1}\}$.  Furthermore, the models can have, besides the density, also
momentum, and temperature given as macroscopic variables.

In principle \eqref{eq:CR_LBM} is easily generalized to a problem with
$N_v$ velocoties. The matrix $M$ --- the matrix that transforms the
distribution functions into velocity moments --- is now a transposed
Vandermonde matrix
\begin{equation*}
M = \begin{pmatrix}
1& 1& \ldots & 1 & 1\\
v_0 & v_1 & \ldots & v_{q-2} & v_{q-1}\\
v_0^2 & v_1^2 & \ldots & v_{q-2}^2 & v_{q-1}^2\\
& \vdots &  & \vdots & \\
v_0^{q-1} & v_1^{q-1} & \ldots & v_{q-2}^{q-1} & v_{q-1}^{q-1}\\
\end{pmatrix},
\end{equation*}
with $q=N_v$.
Define $M^0$ as 
\begin{equation*}
M^0 = \begin{pmatrix}
1& 1& \ldots & 1 & 1\\
v_0 & v_1 & \ldots & v_{q-2} & v_{q-1}\\
& \vdots &  & \vdots & \\
v_0^{k-1} & v_1^{k-1} & \ldots & v_{q-2}^{k-1} & v_{q-1}^{k-1}\\
0 & 0 & \ldots & 0 & 0 \\
& \vdots &  & \vdots & \\
0 & 0 & \ldots & 0 & 0 \\
\end{pmatrix},
\end{equation*}
with $k$ the number of conserved macroscopic variables, $1 \leq k <
N_v$.  $M^0$ represents the part of matrix $M$ that produces the lower
order velocity moments which should be conserved in the CR
algorithm. The part corresponding to the higher order velocity moments
is replaced with zeros. When $k=3$ we conserve density, momentum, and
temperature. These macroscopic variables are linked to the three
lowest order velocity moments of the distribution functions.

Since $M$ is a transposed Vandermonde matrix and the inverse of this
matrix is required to write down the projection operator in the first
term of equation \eqref{eq:CR_LBM}, we should deal with the inverse of such a matrix.

A lot of literature elaborates on this problem in the context of
polynomial interpolation. However, in realistic applications more than
hundred velocity directions need to be taken into account. Even with
known techniques, like using the Lagrange polynomials for inverting
the Vandermonde matrix \cite{turner} or by defining the moments in
Chebyshev polynomials \cite{Abramowitz}, it is still impossible to
compute the inverse with lots of velocity directions.  Similarly, the
method discussed by Golub and Van Loan in \cite{golub} to solve a
Vandermonde system by comparing it to polynomial interpolation and
considering the Newton representation of the interpolating polynomial
has analogous issues.

These issues are further outlined below. When $M^{-1}$ is not
computed correctly, the conservation of the given macroscopic
variables cannot be guaranteed. This is demonstrated in Example
\ref{ex_vandermonde} that shows that $P$ is not a projection operator
when $M^{-1}$ is computed incorrectly. We deal with
these issues in the remainder of this section.

\begin{example}[Eigenvalues of $P:=I-M^{-1}M^0$]\label{ex_vandermonde}
This example shows one of the major issues of inverting matrix $M$ to obtain projection operator $P$ in the CR algorithm.
Since $P$ is  a projection operator  the eigenvalues should be either zero or one.
Figure \ref{eigenvalues_P} contains the computed eigenvalues of this operator with $N_v = 56$, $v_\text{min} = -9.9875\cdot 10^3$ m/s, $v_\text{max}=9.9875\cdot 10^3$ m/s, and $\Delta v = (v_\text{max}-v_\text{min})/N_v$. 
This figure shows that we lose the property of the eigenvalues being either zero or one.
\begin{figure}[!htop]
\begin{center}
\includegraphics[width = 0.5\textwidth]{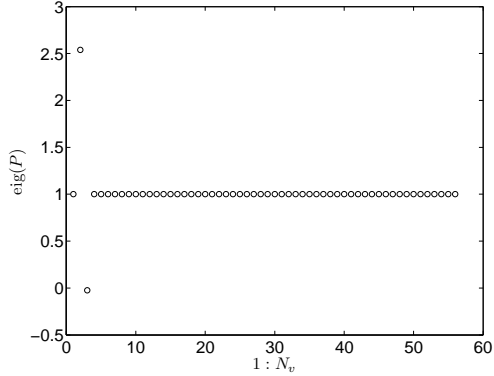}
\caption{Computed eigenvalues of operator $P=I-M^{-1}M^0$ with $N_v = 56$, $v_\text{min} = -9.9875\cdot 10^3$ m/s, $v_\text{max}=9.9875\cdot 10^3$ m/s, and $\Delta v = (v_\text{max}-v_\text{min})/N_v$. $P$ should be a projection operator thus we expect the eigenvalues to be either zero or one. This figure shows that we lose this property when using the inverse of matrix $M$ in the numerical experiment. 
\label{eigenvalues_P}}
\end{center}
\end{figure}
\end{example}

Example \ref{ex_vandermonde} illustrates the difficulties that
originate from inverting matrix $M$. It is clear that the loss of
accuracy leads to eigenvalues that deviate from zero or one. This
has important consequences for the Constrained Runs algorithm since
the conservation of the lower order velocity moments is required. And
this is not longer the case with incorrect projection operators.

Example \ref{ex_chebyshev} incorporates the Chebyshev polynomials to compute $M^{-1}$. Similarly, we lose the property of eigenvalues being either zero or one. 
\begin{example}[Eigenvalues of $P:=I-M^{-1}M^0$ by defining the moments in Chebyshev polynomials]\label{ex_chebyshev}
This example checks the property that $P$ should be a projection operator when $P$ is based on defining the velocity moments in Chebyshev polynomials.
Figure \ref{eigenvalues_PChebyshev} contains the computed eigenvalues of this operator with $N_v = 56$, $v_\text{min} = -9.9875\cdot 10^3$ m/s, $v_\text{max}=9.9875\cdot 10^3$ m/s, and $\Delta v = (v_\text{max}-v_\text{min})/N_v$. 
This figure shows that we lose the property of the eigenvalues being either zero or one.
\begin{figure}[!htop]
\begin{center}
\includegraphics[width = 0.5\textwidth]{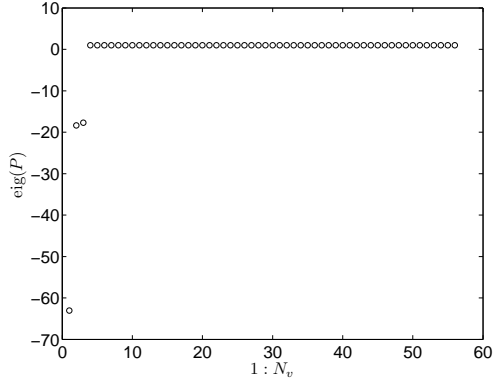}
\caption{Computed eigenvalues of operator $P=I-M^{-1}M^0$ with $N_v = 56$, $v_\text{min} = -9.9875\cdot 10^3$  m/s, $v_\text{max}=9.9875\cdot 10^3$ m/s, and $\Delta v = (v_\text{max}-v_\text{min})/N_v$. $P$ should be a projection operator thus we expect the eigenvalues to be either zero or one. This figure shows that we lose this property when using the inverse of matrix $M$ in the numerical experiment when $M$ is based on defining the velocity moments in Chebyshev polynomials. 
\label{eigenvalues_PChebyshev}}
\end{center}
\end{figure}
\end{example}

However, the use of $M^{-1}$ can be circumvented. The projection on the
space of higher order moments can be written as a subtraction of the
components in the direction of the lower order moments. This then only
requires orthogonal vectors that span the space of lower order moments.

Let us look at the QR factorization of ${M^0}^T(:,1\!:\!k)$ where $k$
is the number of conserved macroscopic variables:
\begin{equation*}
{M^0}^T(:,1\!:\!k) = QR,
\end{equation*}
where $Q$ contains $k$ orthogonal columns and $R$ is a $k$ by $k$
upper triangular matrix.  When density, momentum, and energy are
conserved in a one-dimensional problem $R$ is a 3$\times$3 matrix and $Q$ contains three columns.
Let us define $\hat{M}^0 :=  [ Q^T; 0]$.  We now have that {$M^0(1\!:\!k,:) = R^T \hat{M}^0(1\!:\!k,:)$}.
The space spanned by the columns of $\hat{M}^0$ now spans the same space
as the original low order moments.

The CR iteration $\vect{f}^{(k+1)} = \hat{P} \vect{f}^{(k)\text{pre}} + \hat{M}^0
\vect{f}^0$, where $\hat{P}$ projects on the space of the higher order
moments, can now be written as
\begin{equation}\label{CR_hat}
  \vect{f}^{(k+1)} = (1 - \mbox{$\hat{M}^0$}^T \hat{M}^0) \vect{f}^{(k)\text{pre}} + \hat{M}^0 \vect{f}^0,
\end{equation}
which avoids the inverse of the moment matrix.  \eqref{CR_hat} can be written because
$\hat{M}^T \hat{M}^0 = \mbox{$\hat{M}^0$}^T(:,1\!:\!k)
\hat{M}^0(1\!:\!k,:)$, where $\hat{M}$ is the matrix that would
have been obtained by a $QR$ factorization of the full moment matrix $M$. 

We will demonstrate that an iteration using $\hat{M}$ will preserve the same moments as the original procedure based on $M$. However, the procedure based on $\hat{M}$ is more stable, which can be observed from the eigenvalues of the projection operator. 
Example \ref{ex_hatP} shows these eigenvalues for a similar numerical experiment as in Example \ref{ex_vandermonde} and Example \ref{ex_chebyshev}.

\begin{example}[Eigenvalues of $\hat{P}:=I-\hat{M}^T\hat{M}^0$]\label{ex_hatP}
The eigenvalues of $\hat{P}$ should be either zero or one since $\hat{P}$ is a projection operator.
Figure \ref{eigenvalues_hatP} shows the computed eigenvalues of this operator with $N_v = 56$, $v_\text{min} = -9.9875\cdot 10^3$ m/s, $v_\text{max}=9.9875\cdot 10^3$ m/s, and $\Delta v = (v_\text{max}-v_\text{min})/N_v$. 
The property of the eigenvalues being either zero or one is fulfilled for this numerical experiment.
\begin{figure}[!htop]
\begin{center}
\includegraphics[width = 0.5\textwidth]{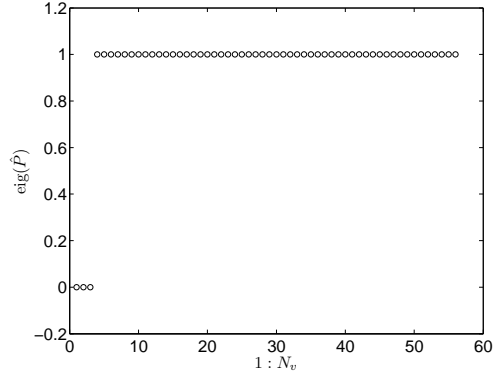}
\caption{Computed eigenvalues of operator $\hat{P}=I-\hat{M}^{T}\hat{M}^0$ with $N_v = 56$, $v_\text{min} = -9.9875\cdot 10^3$ m/s, $v_\text{max}=9.9875\cdot 10^3$ m/s, and $\Delta v = (v_\text{max}-v_\text{min})/N_v$. $\hat{P}$ should be a projection operator thus we expect the eigenvalues to be either zero or one. This figure shows that operator $\hat{P}$ fulfills this property for the numerical experiment.  \label{eigenvalues_hatP}}
\end{center}
\end{figure}
\end{example}

Example \ref{ex_hatP} shows that this alternative setting fulfills the property of eigenvalues being zero or one for projection operator $\hat{P}$.
We check the effect of this redefinition --- using $\hat{P}$ instead of $P$ --- on the CR algorithm.

Now, we check whether this algorithm conserves the same lower order velocity moments, the $k$ conserved macroscopic variables
\begin{align*}
M^0(1\!:\!k,:)\vect{f}^{(k)} &= M^0(1\!:\!k,:)\left(\hat{P}\vect{f}^{(k)\text{pre}}+(I-\hat{P})\vect{f}^0\right), \\
&= M^0(1\!:\!k,:)\hat{P}\vect{f}^{(k)\text{pre}} + M^0(1\!:\!k,:)\vect{f}^0 - M^0(1\!:\!k,:)\hat{P}\vect{f}^0, \\
&= R^T\hat{M}^0(1\!:\!k,:)\hat{P}\vect{f}^{(k)\text{pre}} + M^0(1\!:\!k,:)\vect{f}^0 - R^T\hat{M}^0(1\!:\!k,:)\hat{P}\vect{f}^0.
\end{align*}
The first and last term of the right-hand side vanish because of the definition of the projection space $\hat{W}$ of operator $\hat{P}$. It follows that
\begin{equation*}
M^0(1\!:\!k,:)\vect{f}^{(k)} = M^0(1\!:\!k,:)\vect{f}^0.
\end{equation*}
This confirms the correspondence between the lower order moments in
the CR algorithm and those of $\vect{f}^0$, the initial distribution
function.


\section{Numerical results} \label{numerical_results} We illustrate
the proposed method with the help of a model problem coming
from laser ablation. Here a Boltzmann model is studied with a left
boundary condition that models the evaporation of material from a
heated surface, while the right boundary models the ambient gas. We
refer to \cite{gusarov} for a detailed discussion of the physics and
the Boltzmann model. 
This section contains numerical
results which test the generalized CR algorithm in a setting of
restriction and lifting.

\begin{example}[One-dimensional Helium problem]\label{example_helium}
As a model problem, we consider the one-dimensional laser ablation problem with Helium as a background gas. 
The ambient gas parameters $p_a,T_a,n_a,\rho_a$, and $u_a$ represent the  pressure, 
temperature, number density, mass density, and average flow velocity while 
$p_s,T_s,n_s,\rho_s$, and $u_s$ are the surface parameters. These parameters, presented below in \eqref{param_surface_ambient}, are used for initialization and to obtain boundary conditions. These are similar to the parameters presented in \cite{gusarov}. The boundary conditions at the surface are based on the equilibrium distribution of the surface parameters while the outer boundary is placed far enough from the interface such that the vapor is in equilibrium there.
\begin{align} \label{param_surface_ambient}
&  p_a=0.10132500\cdot 10^6 \: Pa, \quad T_a=0.30000785\cdot 10^3 \: K, \quad n_a=\frac{p_a}{k_B T_a}, \nonumber \\
&  \rho_a=mn_a, \quad u_a=0, \quad  p_s = \frac{p_a}{0.3}, \quad T_s = \frac{T_a}{0.2}, \quad n_s = \frac{p_s}{k_BT_s},\nonumber \\ 
&  \rho_s = mn_s, \quad u_s = 0.
\end{align}
We focus on the finite volume discretization, outlined in section \ref{mesoscopic}, with discretization parameters listed below in \eqref{parameters_30000}.
\begin{align} \label{parameters_30000}
&  N = 1600, \quad N_v = 56, \quad u_0=\sqrt{\frac{2k_B T_s}{m}}, \quad v_\text{min} = -4u_0, \quad v_\text{max} = 4u_0, \nonumber\\
&  \Delta v = \frac{v_\text{max}-v_\text{min}}{N_v}, \quad \vect{v} = \left[v_\text{min}+ \frac{\Delta v}{2}:\Delta v:v_\text{max}-\frac{\Delta v}{2}\right], \quad \lambda = \frac{1}{\sqrt{2}\pi d^2 n_s}, \nonumber\\
&  L = 30000\lambda, \quad h = \frac{L}{N}, \quad \vect{x} = \left[\frac{h}{2}:h:\frac{h}{2}+(N-1)h\right], \nonumber\\
&   \mu_\text{ref} = 0.19\cdot 10^{-4} \: Pa \cdot s, \quad T_\text{ref} = 273.15 \: K, \quad \omega(\vect{x},t) = \frac{\rho(\vect{x},t) k_B/m T(\vect{x},t)}{\mu_\text{ref}\left(T(\vect{x},t)/{T_\text{ref}}\right)^{0.66}}, \nonumber \\
&  \Delta t =  0.9\frac{ 1}{\text{max}(\vect{v})/h+\text{max}(\omega(\vect{x},0))},
\end{align}
with $d$ the gas-kinetic molecular diameter and $\lambda$ the mean free path. The time step $\Delta t$ is defined in such a way for stability reasons \cite{mieussens}.
\end{example}

Since the length of the domain determines the discretization parameters, we can also consider
\begin{align} \label{parameters_30}
L = 30\lambda,
\end{align}
which is a valid choice for the length since the thickness of the nonequilibrium layer is about 10-20 mean free path lengths \cite{gusarov}.
The remaining discretization parameters are the same as those presented in \eqref{parameters_30000}.

These different domain lengths determine the possibility of using the CR algorithm as a lifting operator. This is one of the issues discussed in section \ref{sec_conv_CR}, which is related to the convergence rate of CR.


\subsection{Comparison of $\boldsymbol{P}$ and $\boldsymbol{\hat{P}}$ in the CR algorithm}\label{sec_conv_CR}
\begin{figure}[!htop]
\begin{center}
\includegraphics[width = 0.5\textwidth]{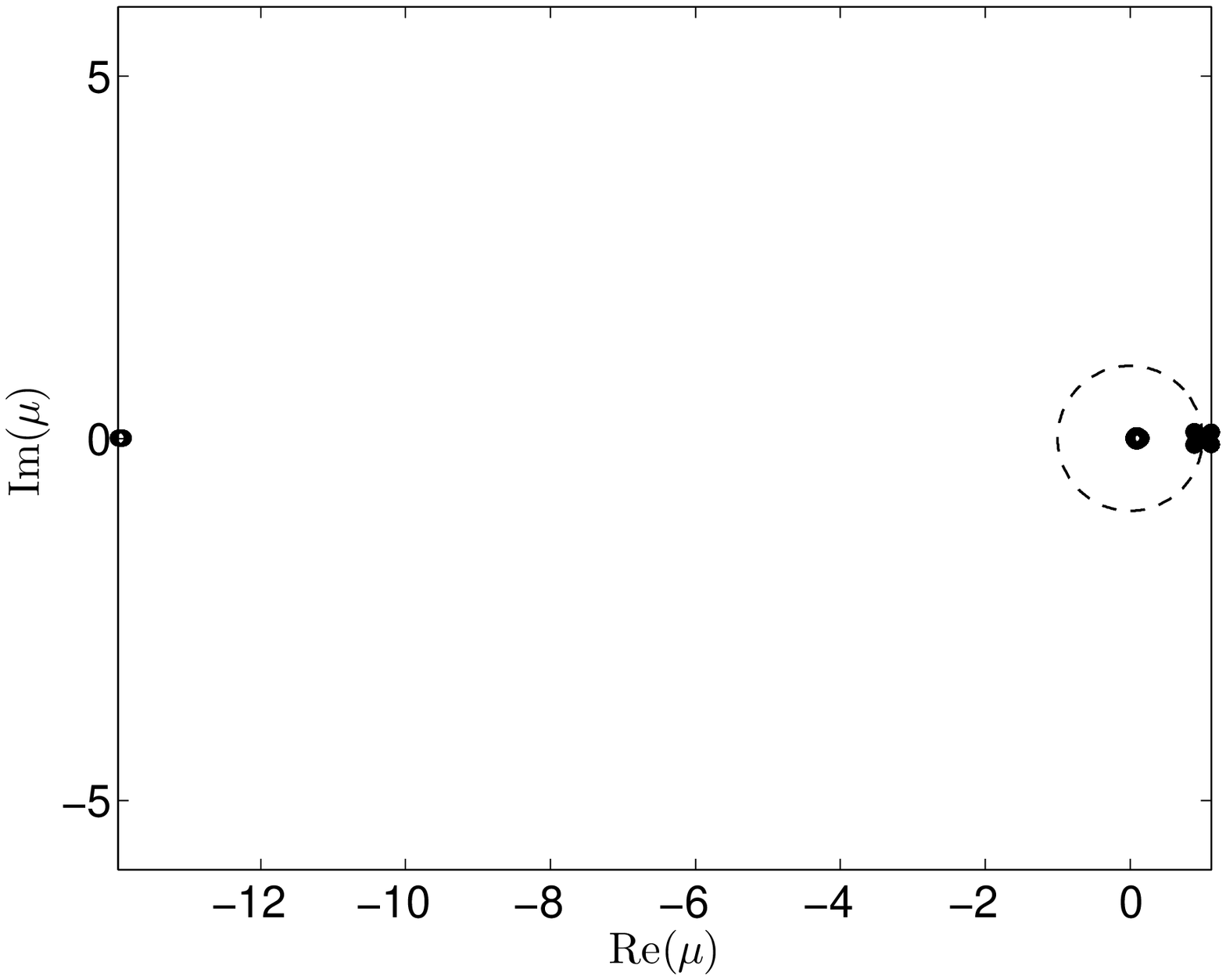}\includegraphics[width=0.5\textwidth]{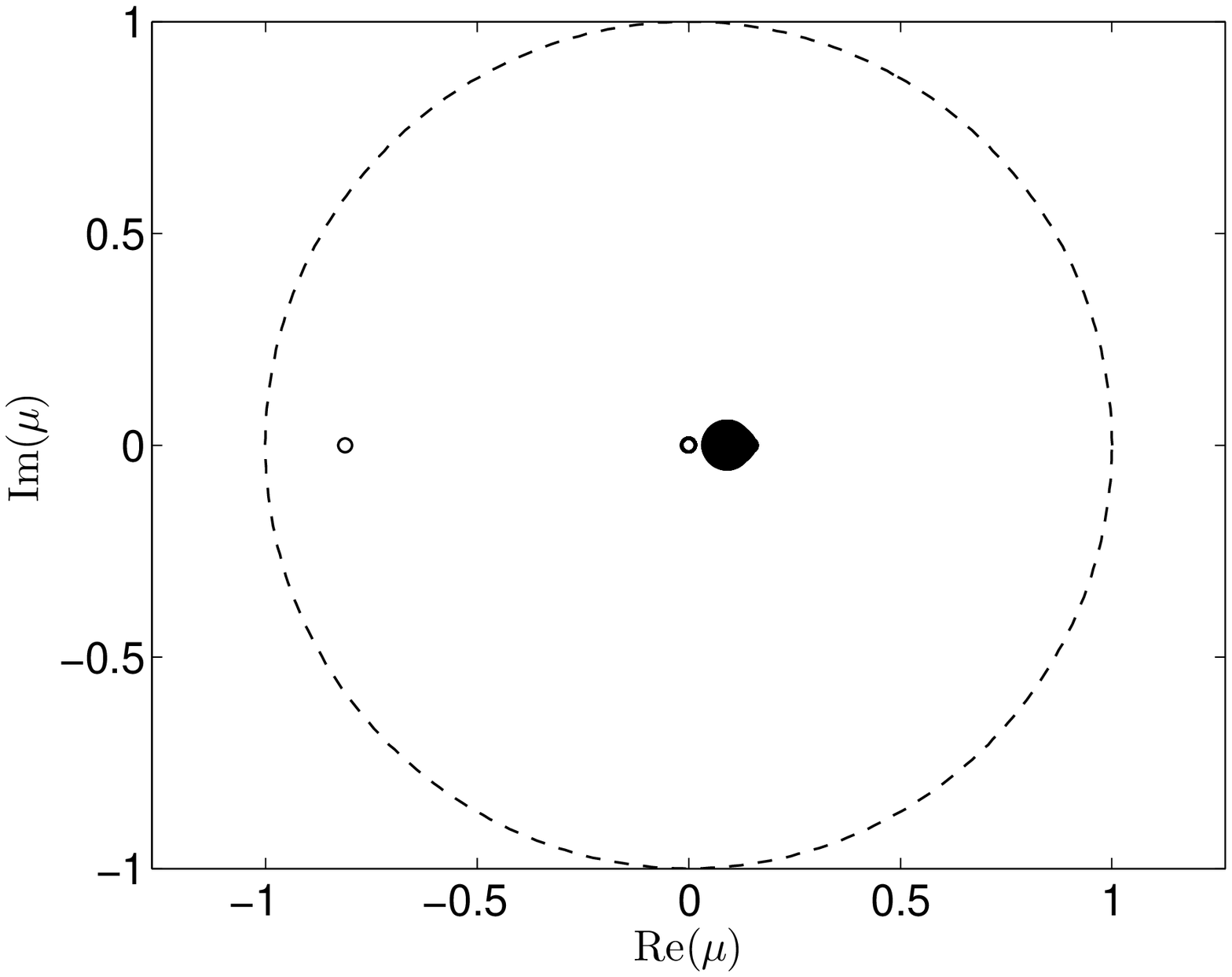}
\caption{The spectrum of the Jacobian matrix of the Constrained Runs map with $m=0$. The left figure uses operator $P$ while the right figure is based on projection operator $\hat{P}$. The parameters are listed in \eqref{parameters_30000} of Example \ref{example_helium} with a smaller number of grid points, namely $N=200$, to make the computation of the Jacobian matrix possible. The unit circle, which should contain the eigenvalues to obtain a stable method, is shown in dashed lines. A clear instability occurs in the left figure when the original projection operator $P=1 - M^{-1} M^0$ is used in the CR algorithm. \label{projection}}
\end{center}
\end{figure}

The following illustrates the issues of using the inverse of matrix
$M$ in the projection operator of the Constrained Runs
algorithm. Figure \ref{projection} shows both the spectra of the
Jacobian matrices of the CR map with operators $P$, which uses the
inverse of $M$, and $\hat{P}$, which avoids this inverse, respectively
in the left and right figure.  The spectra correspond to a Constrained
Runs algorithm with $m=0$. The eigenvalues should remain within the
unit circle for the algorithm to be stable. Since we are focusing on a
generalization of Constrained Runs, we are not only taking the density
as a given macroscopic variable but also average flow velocity and
temperature. This example includes 56 discrete velocities while preserving the
three lowest order velocity moments for the Helium problem presented
in Example \ref{example_helium}. The number of grid points is $N=200$
as opposed to the general parameters presented in
\eqref{parameters_30000}. This reduction in the number of variables is
necessary to make the computation of the Jacobian matrix of the CR map
possible.

These figures show a clear stability with projection operator
$\hat{P}$ while an instability occurs when using the CR algorithm with
$P$ due to the difficulties of computing the inverse of $M$.

\begin{figure}[!htop]
\begin{center}
\includegraphics[width = 0.5\textwidth]{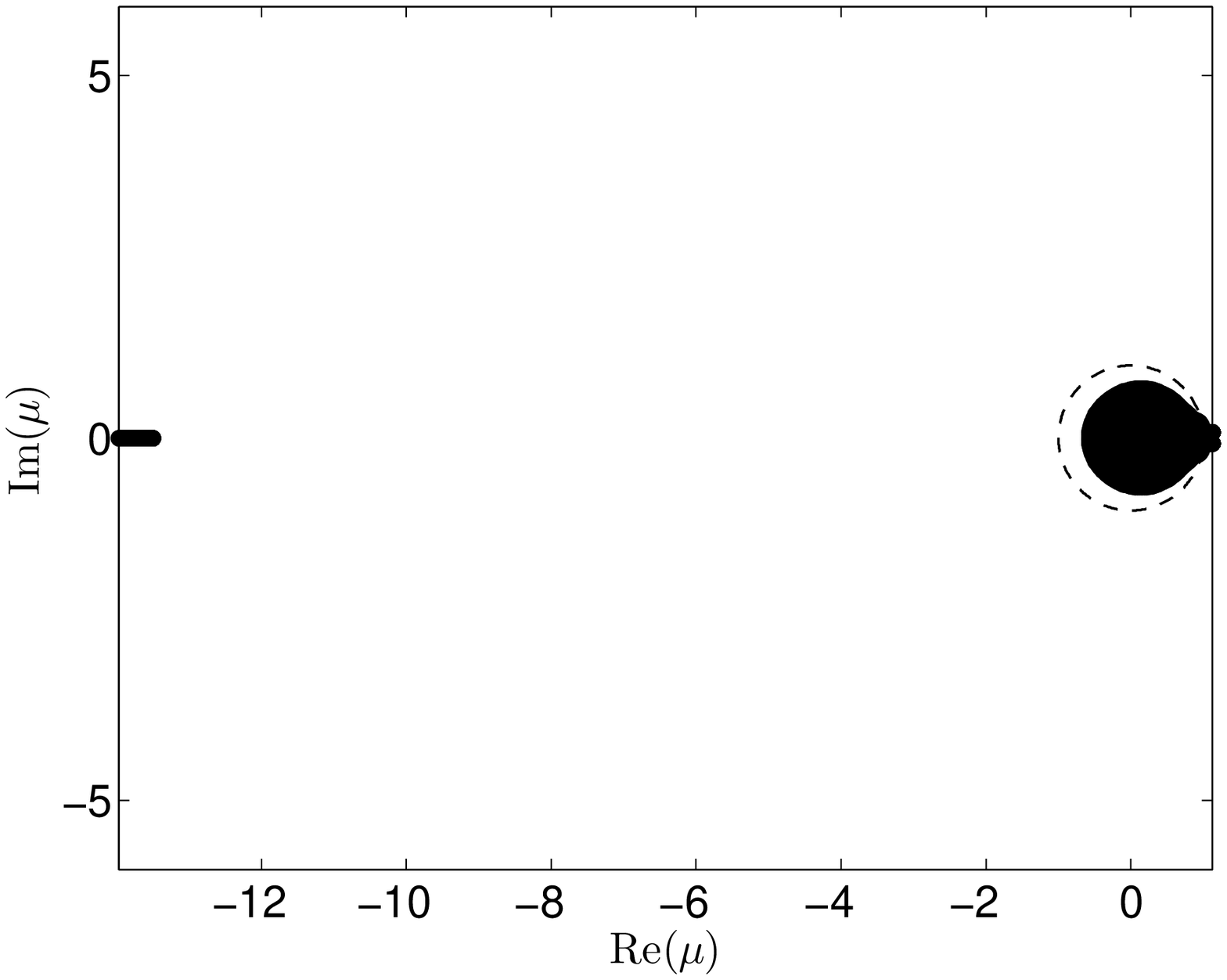}\includegraphics[width=0.5\textwidth]{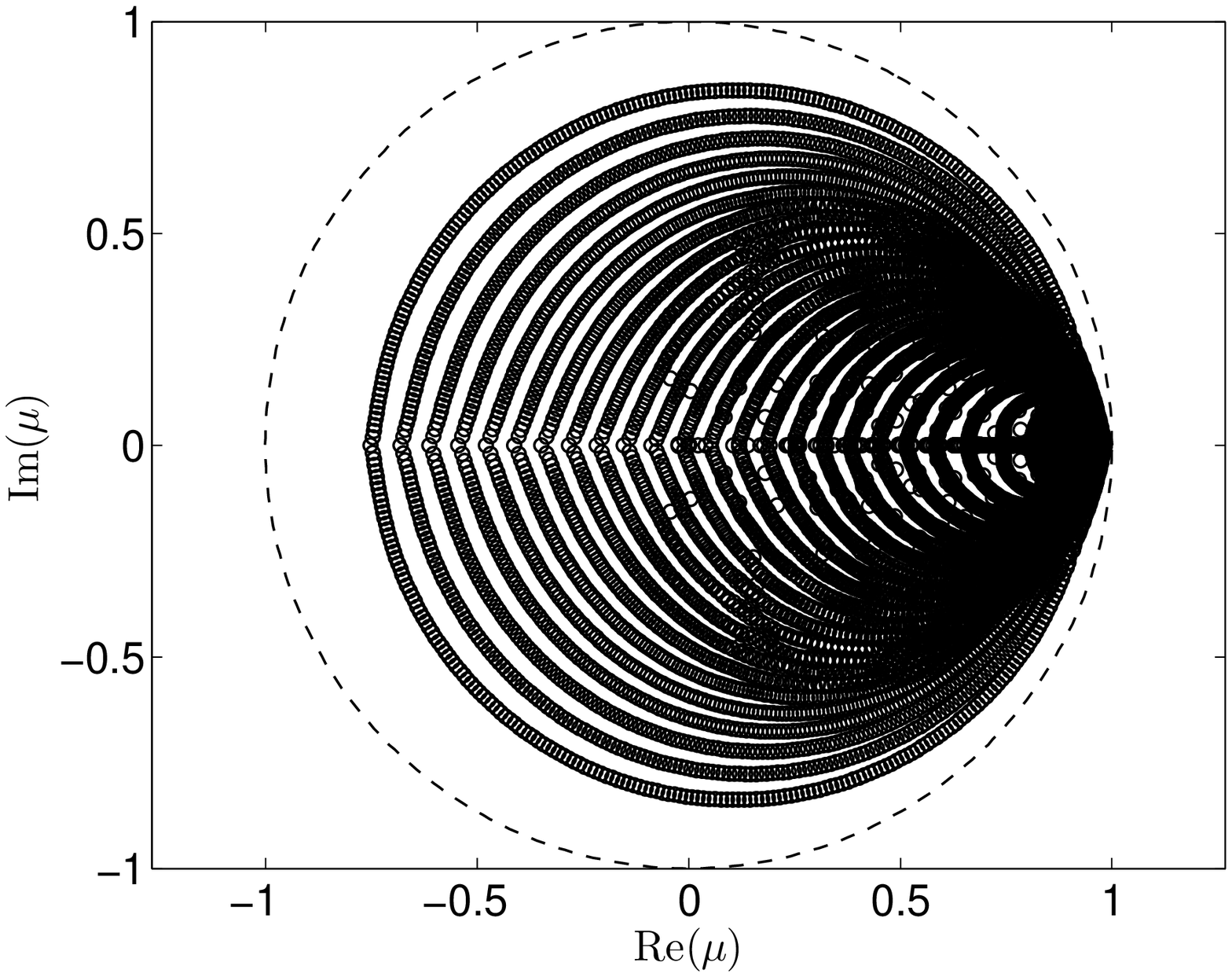}
\caption{ Similar as Figure~\ref{projection} but now with parameters listed in \eqref{parameters_30}, where there ratio of $\Delta x$ to the mean free path is much smaller. \label{projection2}}
\end{center}
\end{figure}

Another issue that is illustrated by the spectrum of the Jacobian matrix of the CR algorithm is the convergence rate of Constrained Runs. Figure \ref{projection2} shows the spectrum of the Helium model problem presented in Example \ref{example_helium} with parameters listed in \eqref{parameters_30}, where the ratio of the grid distance $\Delta x$ to the mean free path is reduced. Again, the CR algorithm with operator $P$ shows a clear instability while $\hat{P}$ stabilizes the method. However, the convergence is slower with these parameters since the spectral radius, or equivalently the asymptotic convergence factor, is larger in Figure \ref{projection2} compared to Figure \ref{projection}.


\subsection{Test Constrained Runs algorithm in Example \ref{example_helium}}
\begin{figure}[!htop]
\begin{center}
\includegraphics[width = 0.5\textwidth]{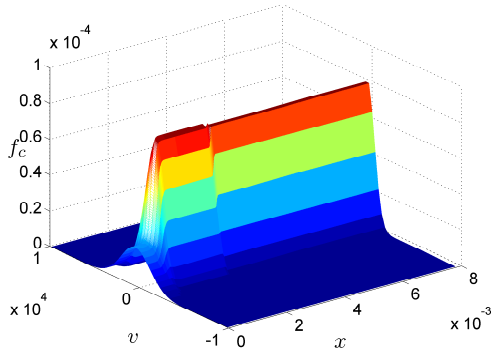}
\caption{Plot of the reference distribution function $\fc$ obtained after 10000 time steps on the initial state with parameters presented in Example \ref{example_helium}. A traveling wave emerges in the domain.  The left boundary condition models the evaporating material in laser ablation. The right boundary is determined by the parameters of the ambient gas. The vertical axis is rescaled with the mass for numerical reasons as described in the text. \label{fig_f_c}}
\end{center}
\end{figure}

This section tests the Constrained Runs algorithm on the model problem presented in Example \ref{example_helium}. 
We perform 10000 Boltzmann time steps on the initial state based on the ambient parameters presented in Example \ref{example_helium} and parameters listed in \eqref{parameters_30000}. The distribution functions are rescaled with the mass $m$ for numerical reasons due too small numbers to perform the numerics on.  This results  in a reference distribution function $\fc$ that  is plotted in Figure \ref{fig_f_c}. As can be seen, a traveling wave emerges in the domain. The corresponding equilibrium distribution  is $\vect{f}^{eq}$.

We test the CR algorithm as a lifting operator that maps density,
average flow velocity, and temperature to distribution functions. The
CR algorithm is combined with Newton's method to ensure stability. It
uses a GMRES algorithm to estimate the Jacobian matrix in Newton's
method. The parameters of the GMRES algorithm are default parameters
in Matlab. The convergence of GMRES might be improved if a preconditioner is used similar to \cite{cai_preconditioner}.

The lifting operator can be evaluated
by restricting the reference distribution function $\fc$ to its
macroscopic variables and lift them back to a distribution function
$\vect{f}$ by using the lifting operator. The resulting $\vect{f}$ is
compared with $\fc$ with the help of the two-norm $\|\vect{f}-\fc\|$,
which is shown in Table \ref{table_CR_helium}. These results can be
compared to $\|\vect{f}^{eq}-\fc\|$ which is equal to $6.4940$e-007.
They are based on Newton's method with a tolerance value of
$1.0$e-010. It might improve with a stricter tolerance.
\begin{table}[!htop]
\caption{The error $\|\vect{f}-\fc\|$ to test the CR algorithm as a lifting operator (combined with Newton's method) for various orders of accuracy. The reference distribution function $\fc$ is obtained for the model problem in Example \ref{example_helium} by performing 10000 Boltzmann time steps on the initial state. 
\label{table_CR_helium}}
\begin{center} \footnotesize
\begin{tabular}{|l | l|}\hline
Order CR & $\|\vect{f}-\fc\|$ \\ \hline
 0 & 1.0428e-006 \\
 1 & 1.6413e-008 \\
 2 & 6.1629e-010 \\
 3 & 4.1965e-010 \\ \hline
\end{tabular}
\end{center}
\end{table}

\begin{figure}[!htop]
\begin{center}
\begin{tabular}{cc}
& $m=0$\\
\includegraphics[width = 0.5\textwidth]{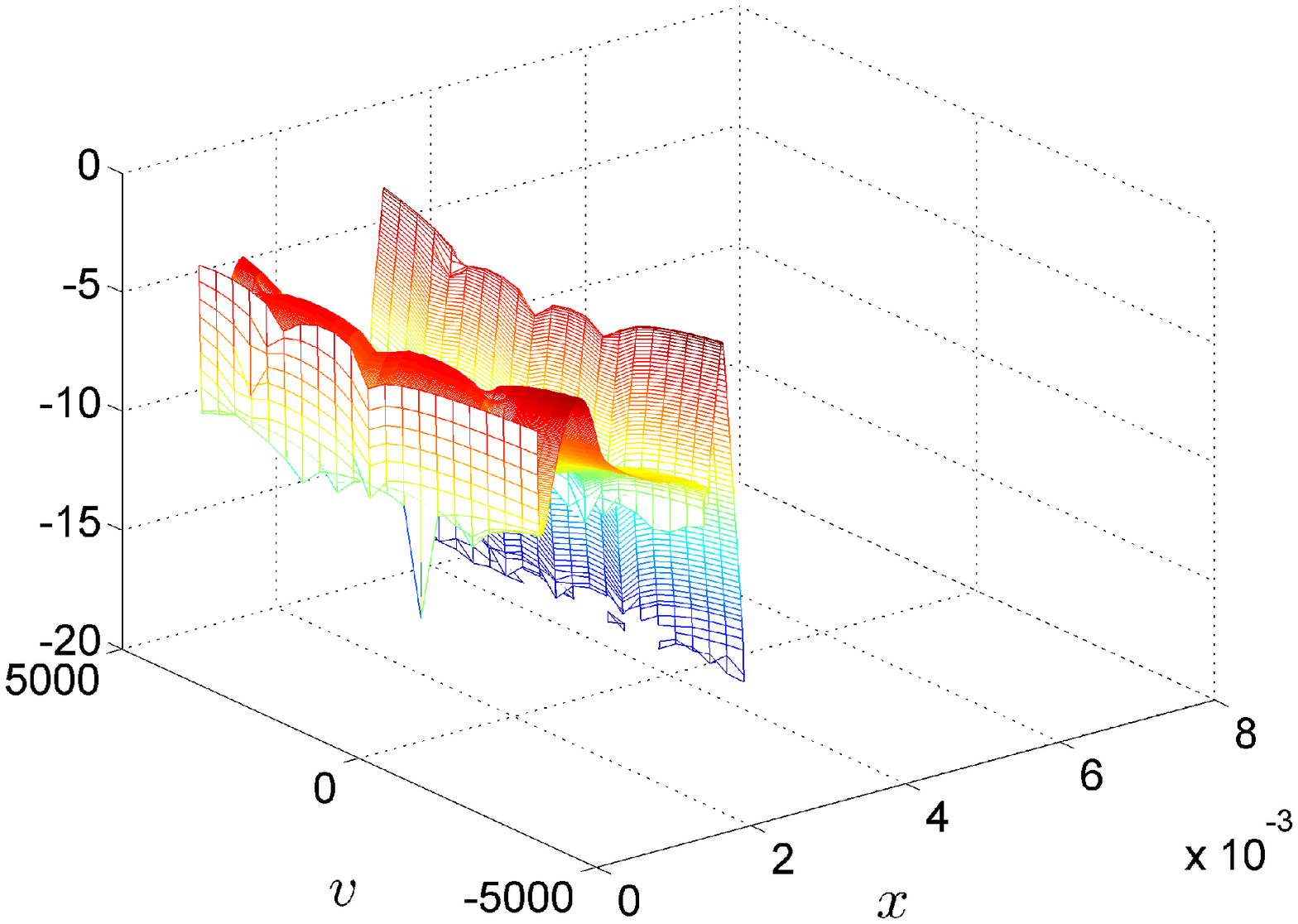}&\includegraphics[width=0.5\textwidth]{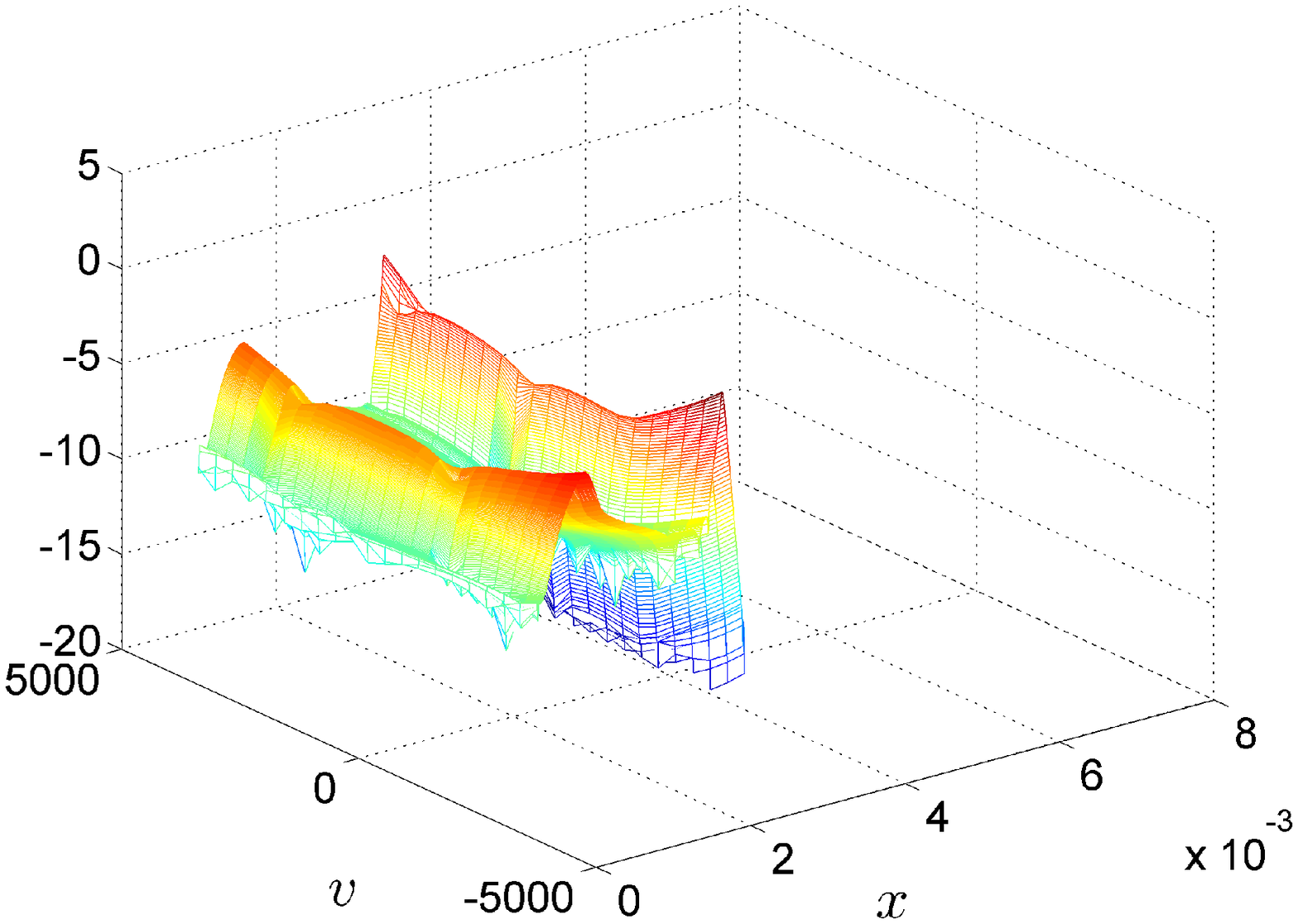} \\
$m=1$&$m=2$\\
\includegraphics[width = 0.5\textwidth]{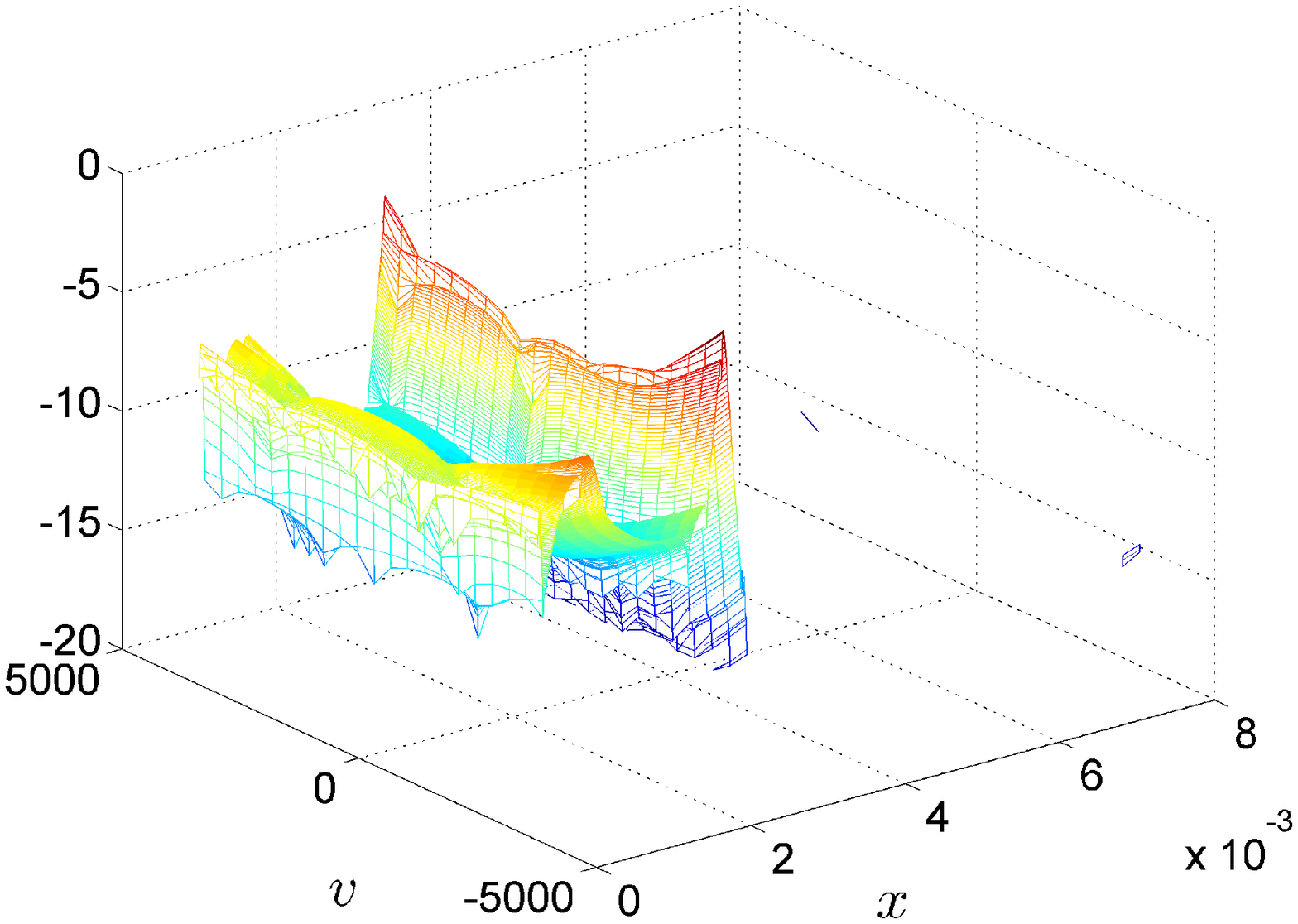}&\includegraphics[width=0.5\textwidth]{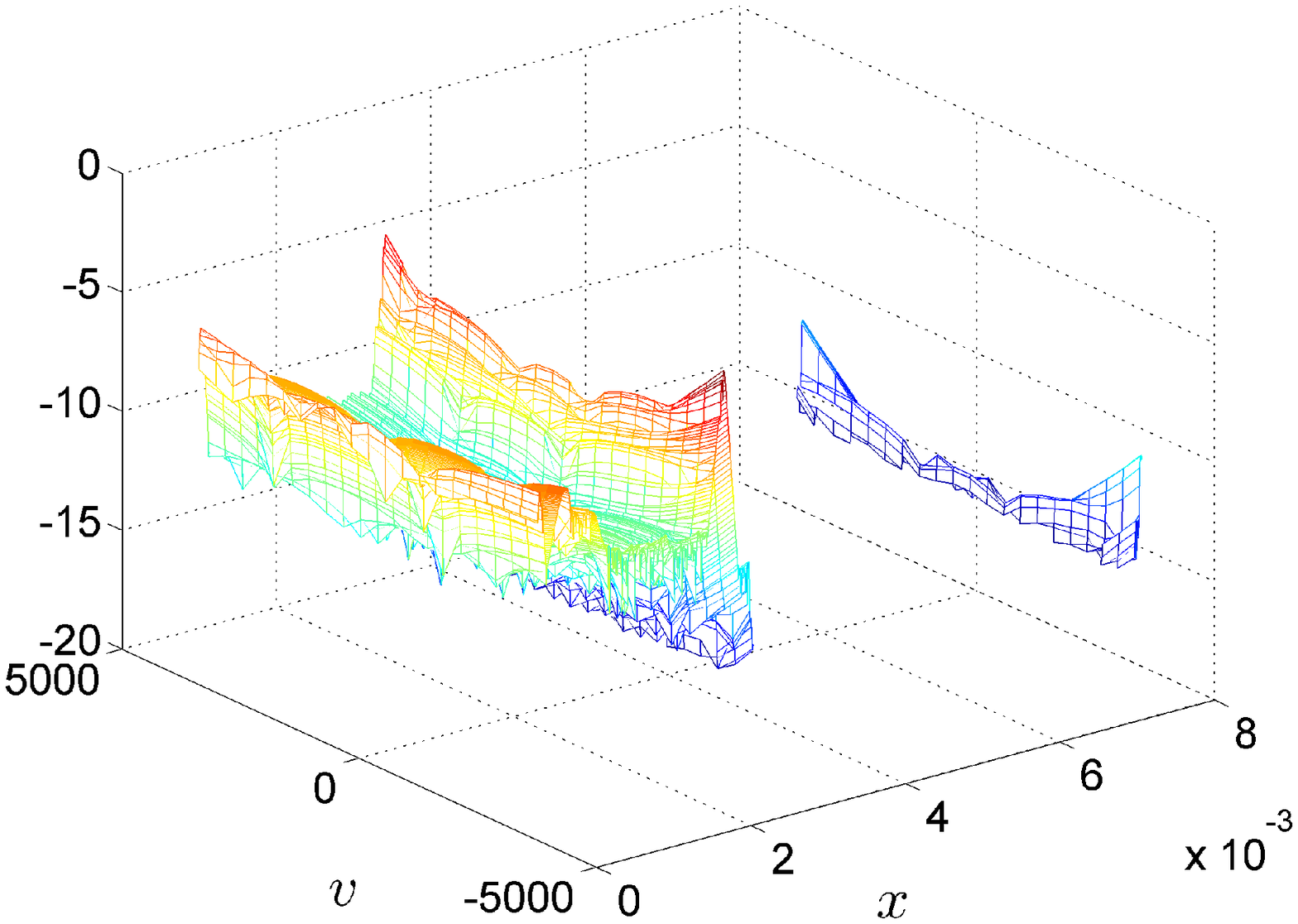} \\
$m=3$&$m=4$\\
\includegraphics[width=0.5\textwidth]{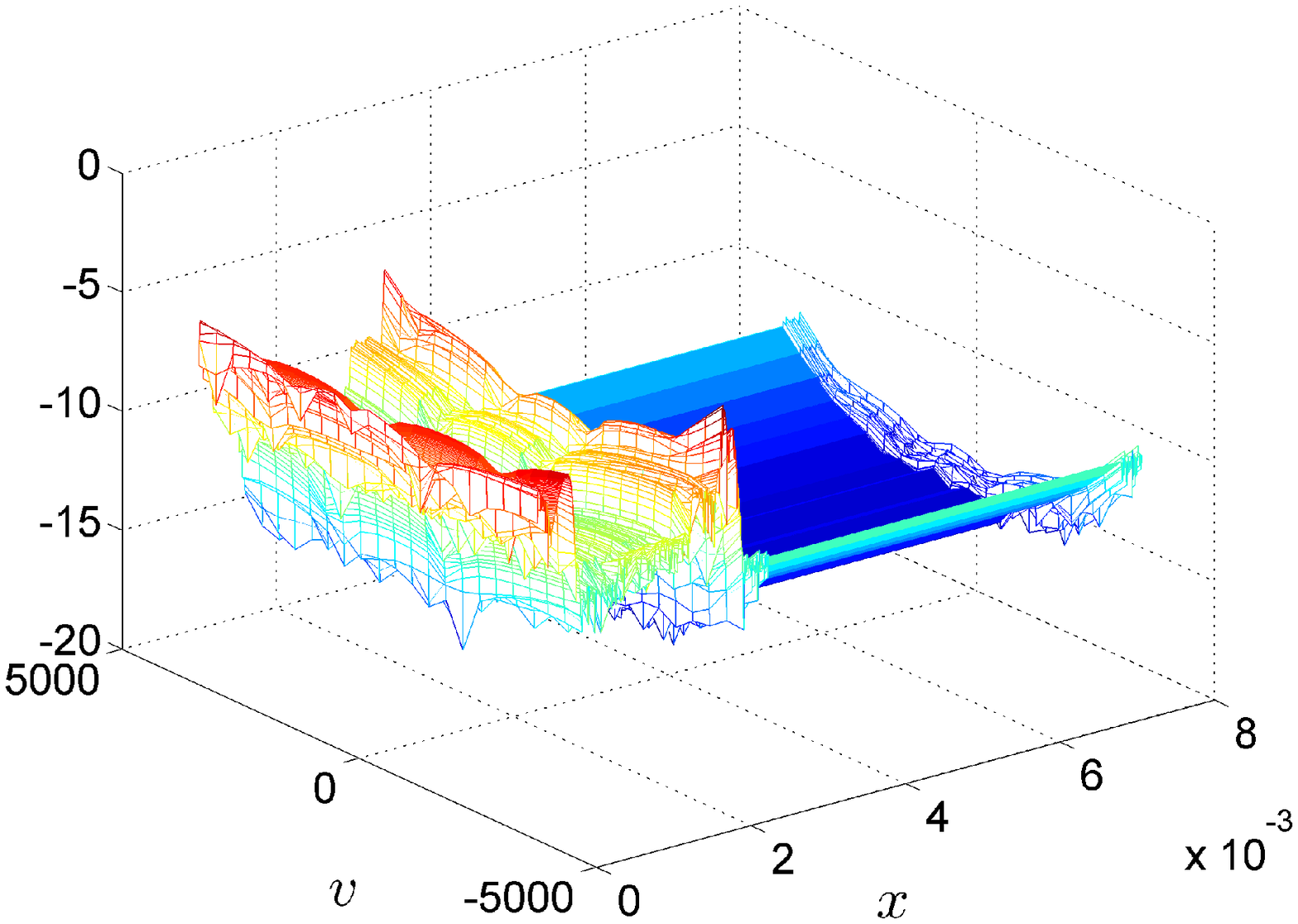}&\includegraphics[width=0.5\textwidth]{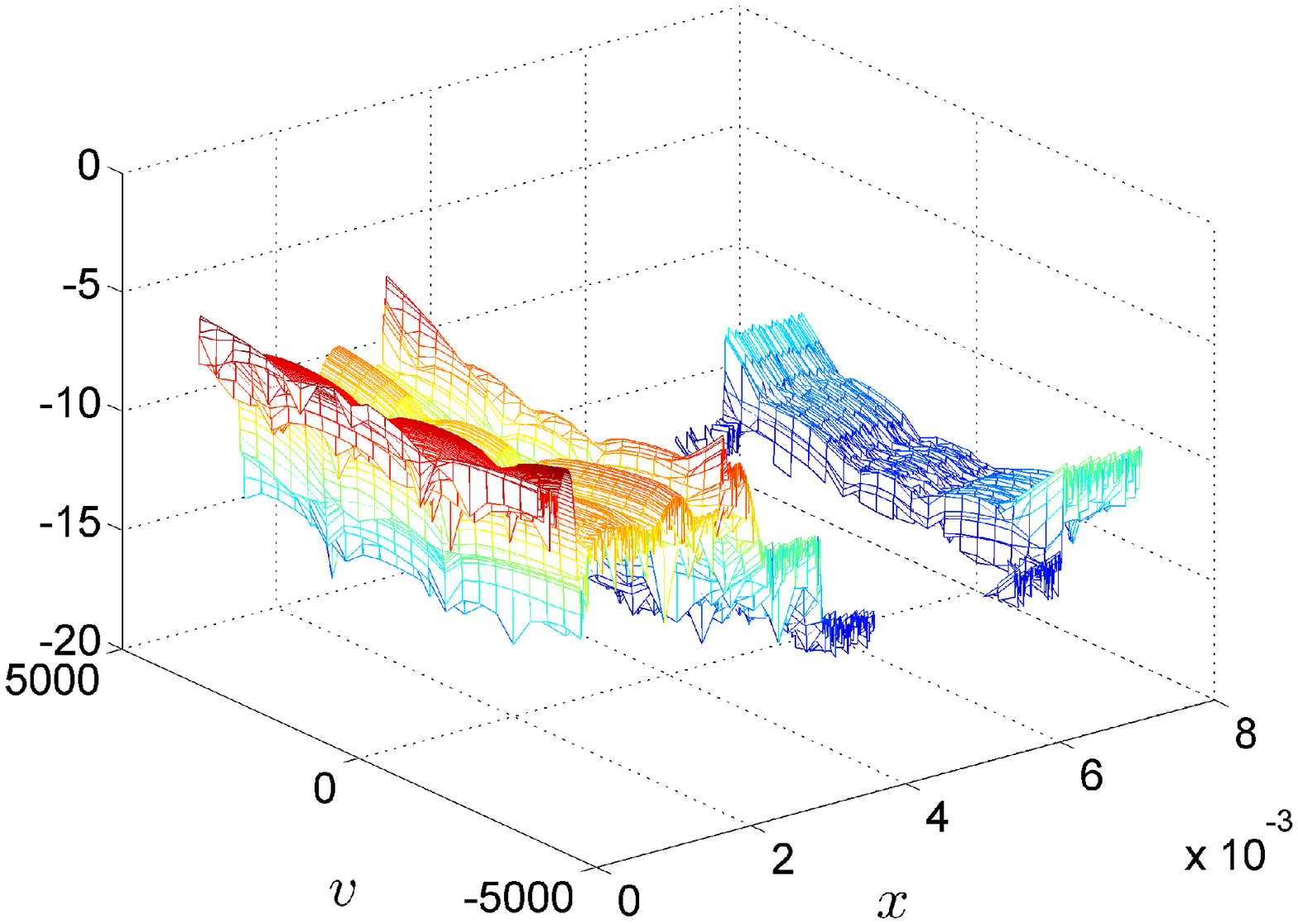}
\end{tabular}
\caption{Log plot of the relative error $\left|\frac{\vect{f}^{eq}-\fc}{\fc} \right|$ (top left) and similarly of $\left|\frac{\vect{f}-\fc}{\fc} \right|$ where $\fc$ represents the reference distribution function after 10000 time steps on the  initial state with parameters presented in Example \ref{example_helium}. $\vect{f}^{eq}$ is the corresponding equilibrium distribution after 10000 steps and $\vect{f}$ the distribution functions based on lifting with the Constrained Runs algorithm of order $m=0$ (top right), $m=1$ (middle left), $m=2$ (middle right), $m=3$ (bottom left), and $m=4$ (bottom right). 
Where the figures show no results (white area in figures), $\left|\vect{f}-\fc\right|$ is exactly equal to zero, which makes it impossible to create the log plot.
\label{fig_CR_helium}}
\end{center}
\end{figure}

Figure \ref{fig_CR_helium} presents a log plot of the relative errors
$\left|\frac{\vect{f}^{eq}-\fc}{\fc} \right|$ (top left) and
$\left|\frac{\vect{f}-\fc}{\fc} \right|$. These results are plotted in
function of spatial grid points $\vect{x}$ and velocities
$\vect{v}$. $\vect{f}$ corresponds to the distribution functions based
on lifting with the Constrained Runs algorithm of order $m=0$ (top
right), $m=1$ (middle left), $m=2$ (middle right), $m=3$ (bottom left),
and $m=4$ (bottom right).  We show a narrow range of velocity
directions since the outer velocities bring the distribution functions
to zero.  Where the figures show no results (white area in figures),
$\left|\vect{f}-\fc\right|$ is exactly equal to zero, which makes it
impossible to create the log plot.

\begin{figure}[!htop]
\begin{center}
\includegraphics[width = 0.5\textwidth]{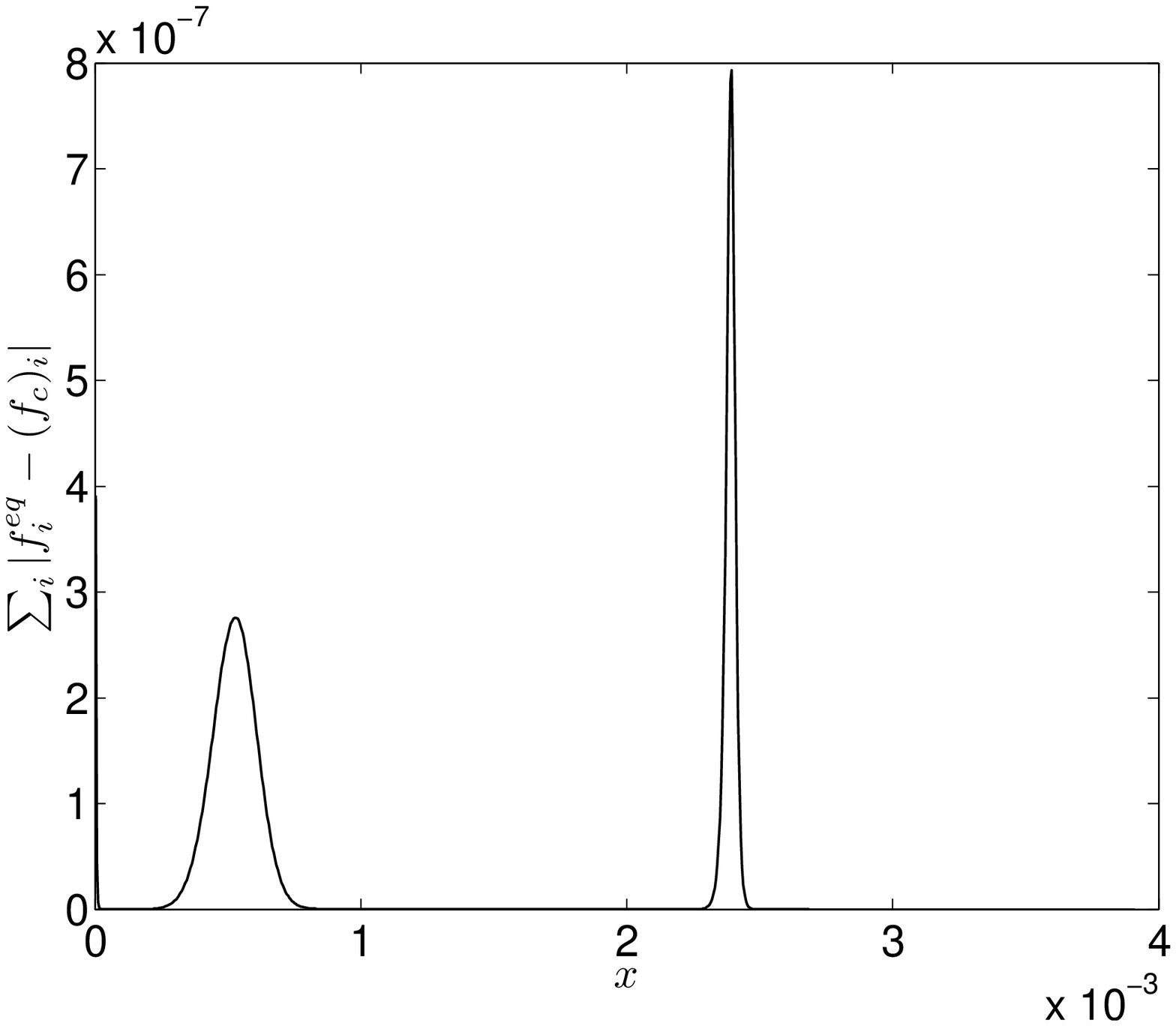}\includegraphics[width = 0.5\textwidth]{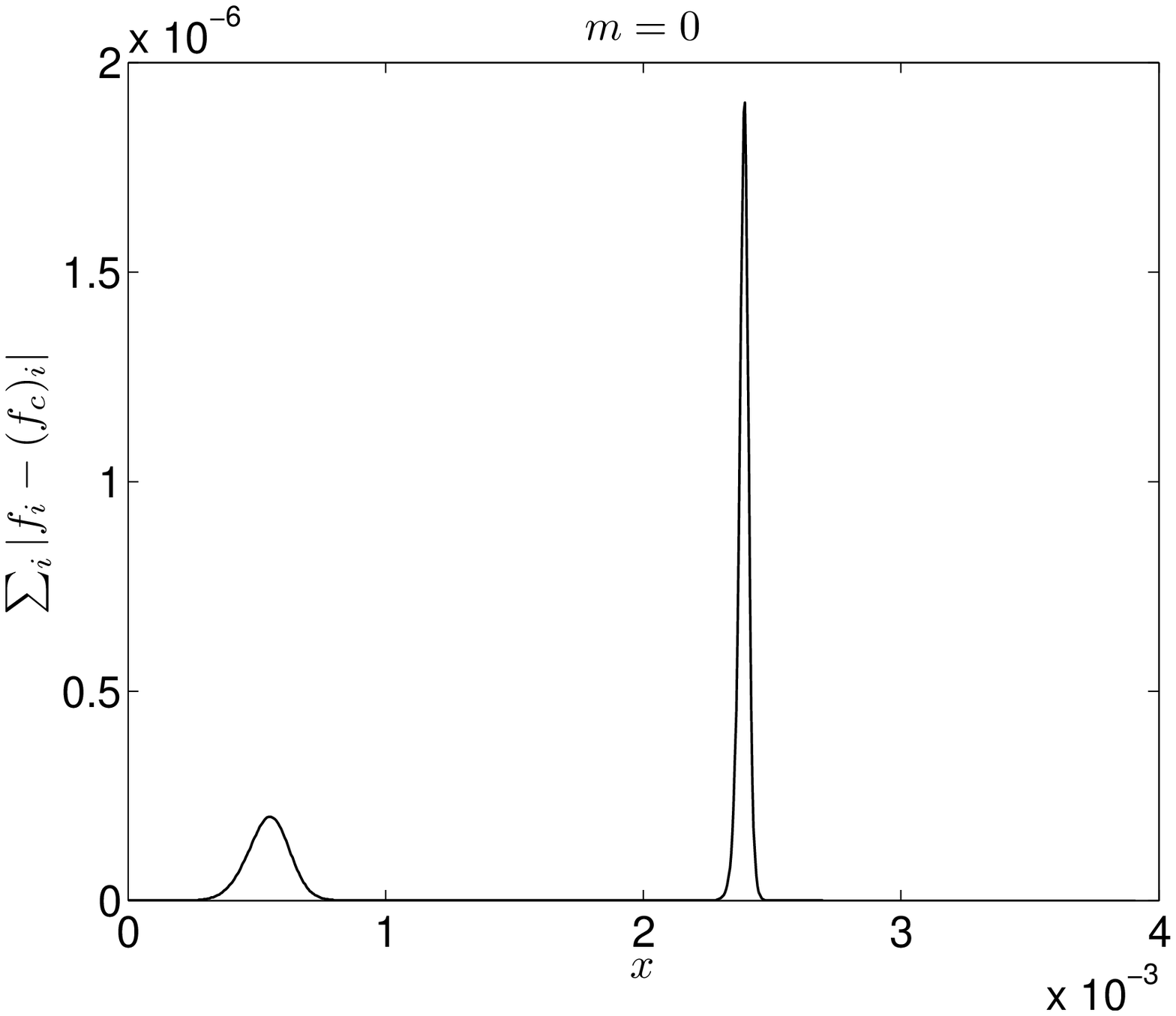}
\includegraphics[width = 0.5\textwidth]{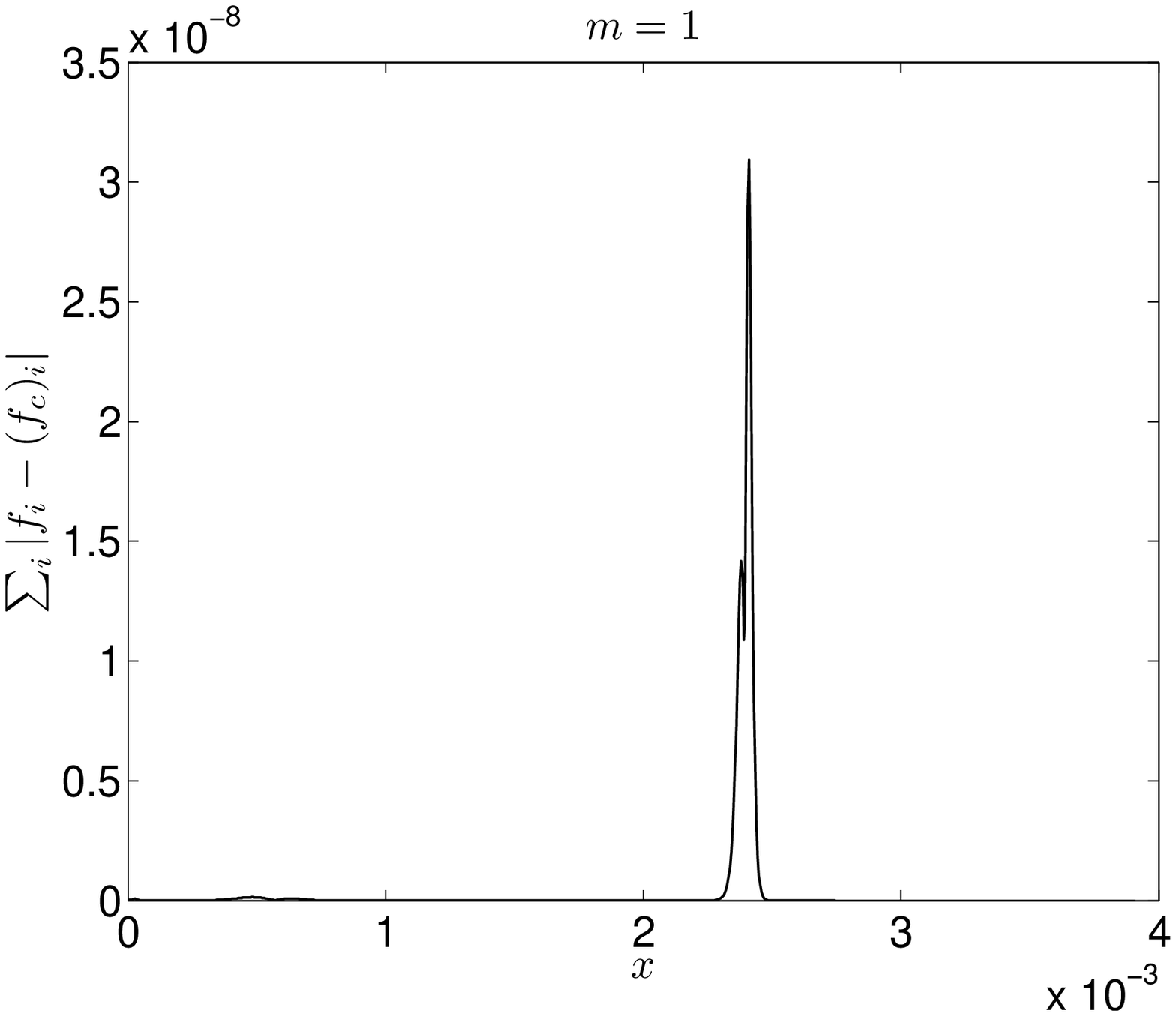}\includegraphics[width = 0.5\textwidth]{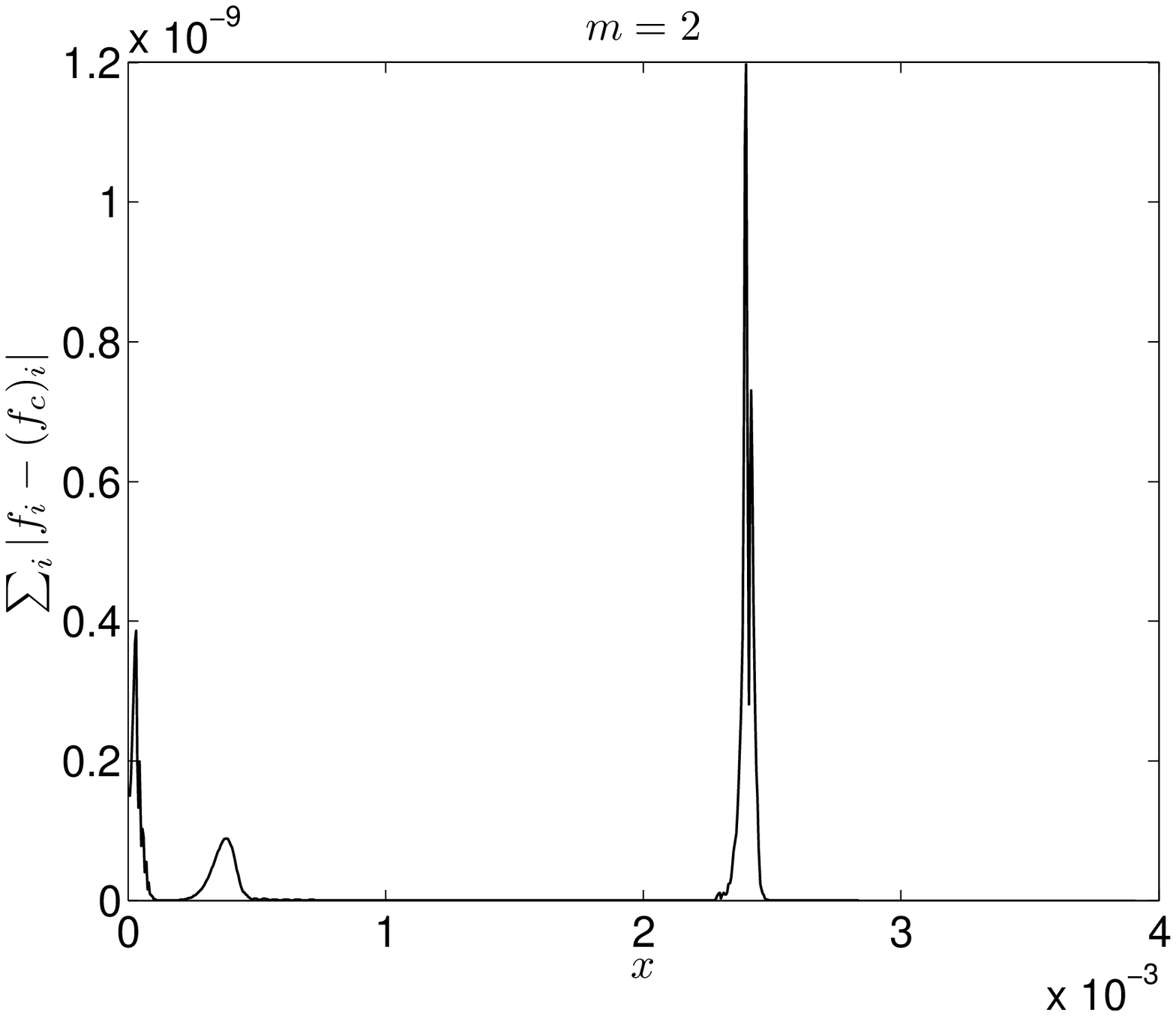}
\includegraphics[width = 0.5\textwidth]{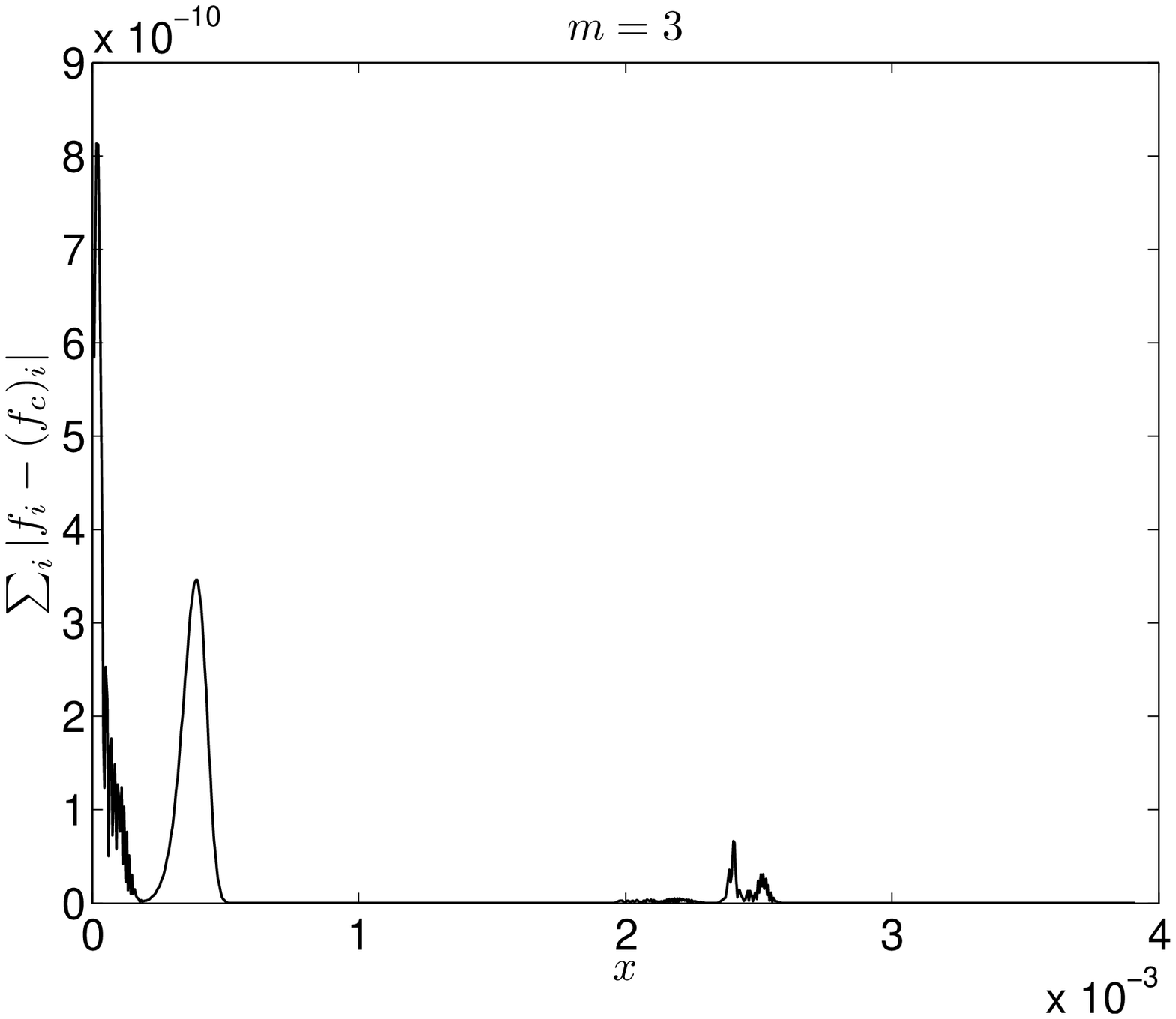}\includegraphics[width = 0.5\textwidth]{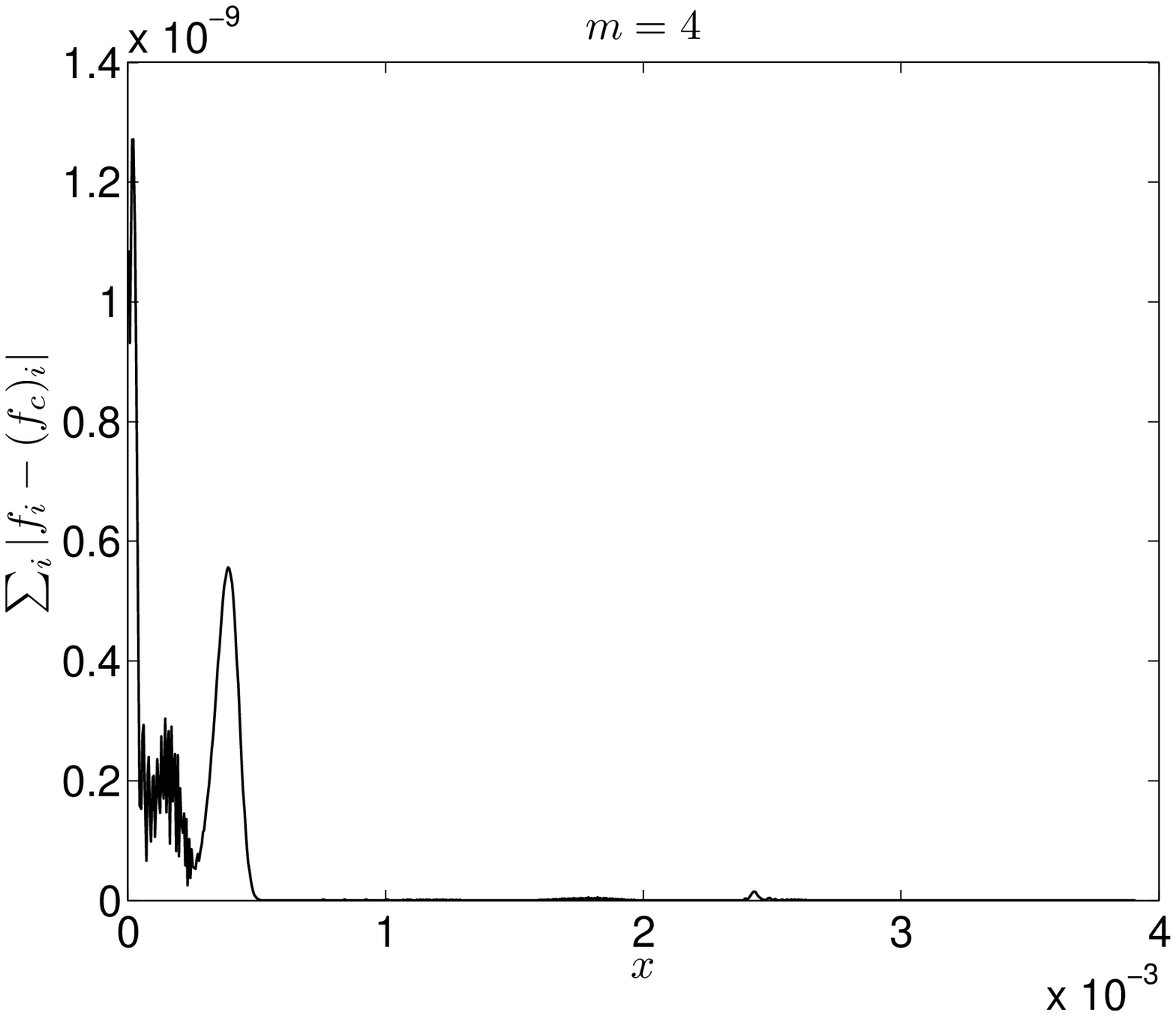}
\caption{Plot of the absolute differences $\sum_i\left|\vect{f}_{\!i}^{eq}-(\fc)_{\!i}\right|$ (top left) and $\sum_i\left|\vect{f}_{\!i}-(\fc)_{\!i}\right|$ where $\fc$ represents the reference distribution function after 10000 time steps on the initial state with parameters presented in Example \ref{example_helium}. $\vect{f}^{eq}$ is the corresponding equilibrium distribution and $\vect{f}$ the distribution functions based on lifting with the Constrained Runs algorithm of order $m=0$ (top right), $m=1$ (middle left), $m=2$ (middle right), $m=3$ (bottom left), and $m=4$ (bottom right). Note the different scales in the figures. These figures are shown to get an idea on the construction of the hybrid domain. \label{abs_diff}}
\end{center}
\end{figure}

Furthermore, we plot $\sum_i \left| \vect{f}_{\!i}^{eq}- (\fc)_{\!i} \right|$ and $\sum_i \left| \vect{f}_{\!i}- (\fc)_{\!i} \right|$ in Figure \ref{abs_diff} to get an idea on the construction of the hybrid domain.

This suggests the use of a Boltzmann model near the traveling wave while a PDE domain can be used outside this part of the domain. Extra care is also needed at the boundaries where the surface parameters determine the state.

\begin{example}[Helium problem including radial velocities]\label{example_helium_radial}
This example considers the laser ablation problem from \cite{gusarov} with Helium as a background gas in a three-dimensional domain described through radial velocities, which represent the axial symmetry.
$p_a,T_a,n_a,\rho_a$, and $u_a$ again correspond to the ambient parameters pressure, temperature, number density, mass density, and average flow velocity while $p_s,T_s,n_s,\rho_s$, and $u_s$ are the surface parameters which are given below. These are similar to the parameters presented in \cite{gusarov}.
\begin{align*}
&  p_a=0.10132500 \cdot 10^6 \: Pa, \quad T_a=0.30000785\cdot 10^3 \: K, \quad n_a=\frac{p_a}{k_B T_a}, \\
&  \rho_a=mn_a, \quad u_a=0, \quad  p_s = \frac{p_a}{0.3}, \quad  T_s = \frac{T_a}{0.2},  \quad n_s = \frac{p_s}{k_BT_s}, \\ 
&  \rho_s = mn_s, \quad u_s = 0.
\end{align*}
The discrete distribution function is defined on a two-dimensional grid with velocities $(v_z)_i=(v_z)_0+i\Delta v_z$, $i\in\{0,\ldots,N_z-1\}$ and $(v_r)_{i'}=(v_r)_0+i'\Delta v_r$, $i'\in\{0,\ldots,N_r-1\}$, specified in the cylindrical domain of the velocity space with $(v_z)_\text{min}<v_z<(v_z)_\text{max}$ and $0<v_r<(v_r)_\text{max}$. $v_r=[(v_x)^2+(v_y)^2]^{1/2}$ is the radial velocity, which takes into account the axial symmetry. The axial and radial velocity increments are $\Delta v_z=((v_z)_\text{max}-(v_z)_\text{min})/N_z$ and $\Delta v_r=(v_r)_\text{max}/N_r$, respectively, $(v_z)_0=(v_z)_\text{min}+\Delta v_z/2$ and $(v_r)_0=\Delta v_r/2$.
The discretization parameters now correspond to
\begin{align} \label{parameters_30000_radial}
&  N = 1600, \quad N_z = 56, \quad N_r = 24, \quad u_0=\sqrt{\frac{2k_B T_s}{m}},  \nonumber\\
&   (v_z)_\text{min} = -4u_0, \quad (v_z)_\text{max} = 4u_0, \quad \Delta v_z = \frac{(v_z)_\text{max}-(v_z)_\text{min}}{N_z}, \nonumber\\
& \vect{v}_z = \left[(v_z)_\text{min}+ \frac{\Delta v_z}{2}:\Delta v_z:(v_z)_\text{max}-\frac{\Delta v_z}{2}\right], \quad (v_r)_\text{min} = 0, \nonumber \\
& (v_r)_\text{max} = 3u_0, \quad \Delta v_r = \frac{(v_r)_\text{max}}{N_r}, \quad \vect{v}_r = \left[\frac{\Delta v_r}{2}:\Delta v_r:(v_r)_\text{max}-\frac{\Delta v_r}{2}\right], \nonumber \\
&  \lambda = \frac{1}{\sqrt{2}\pi d^2 n_s}, \quad L = 30000\lambda, \quad h = \frac{L}{N}, \quad \vect{z} = \left[\frac{h}{2}:h:\frac{h}{2}+(N-1)h\right], \nonumber\\
&  \mu_\text{ref} = 0.19\cdot 10^{-4} \: Pa \cdot s, \quad T_\text{ref} = 273.15 \: K, \quad \omega(\vect{z},t) = \frac{\rho(\vect{z},t) k_B/m T(\vect{z},t)}{\mu_\text{ref}\left(T(\vect{z},t)/T_\text{ref}\right)^{0.66}},  \nonumber \\
&  \Delta t =  0.9\frac{ 1}{\text{max}(\vect{v}_z)/h+\text{max}(\omega(\vect{z},0))}.
\end{align}
\end{example}
Including radial velocities requires a redefinition of the velocity moment matrix $M$. In general, tensors are necessary to describe higher order velocity moments \cite{struchtrup}. The lower order moments, density, average flow velocity, and temperature are given in \eqref{lower_moments}.

\begin{figure}[!htop]
\begin{center}
\includegraphics[width = 0.5\textwidth]{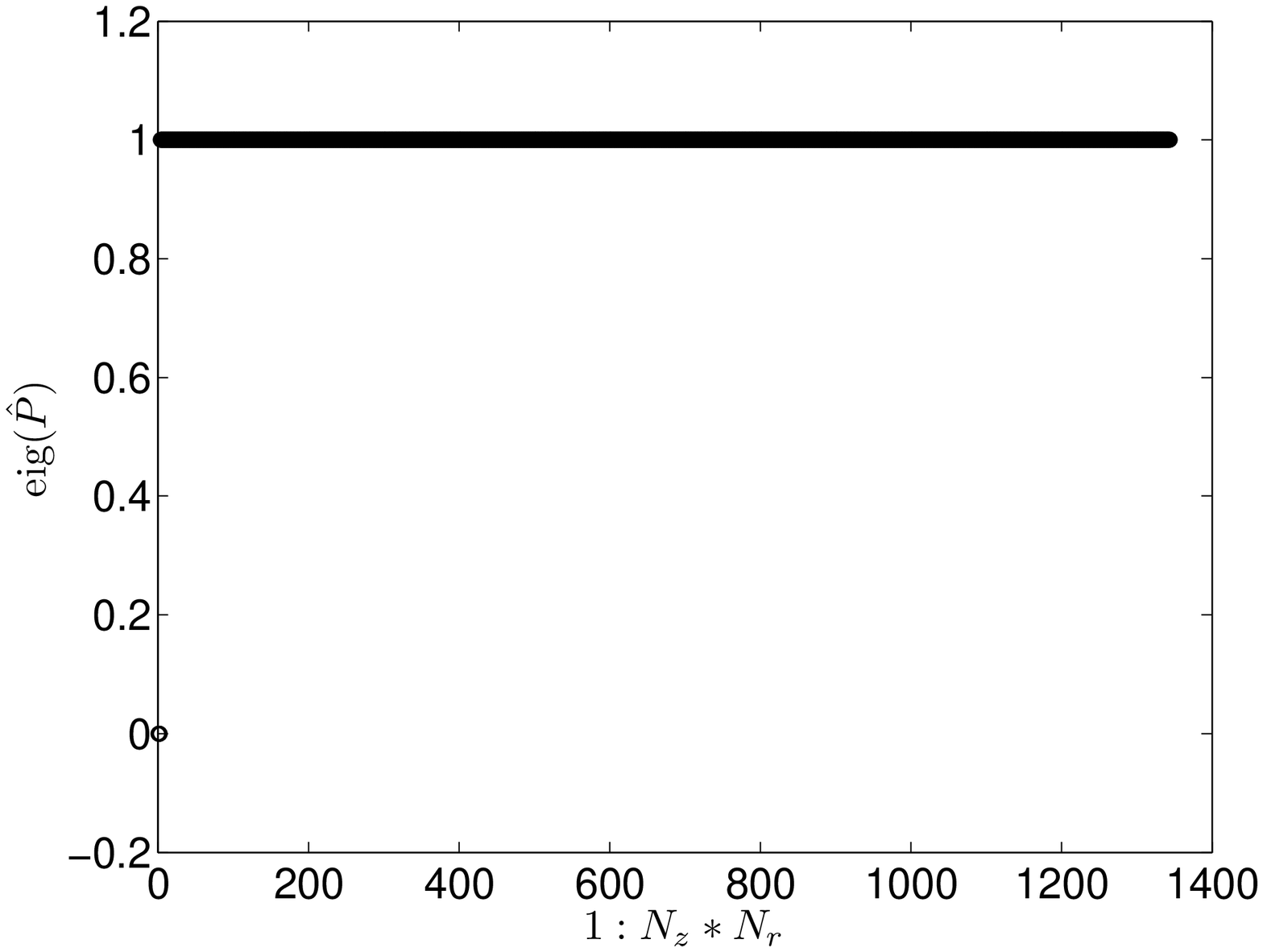}\includegraphics[width = 0.5\textwidth]{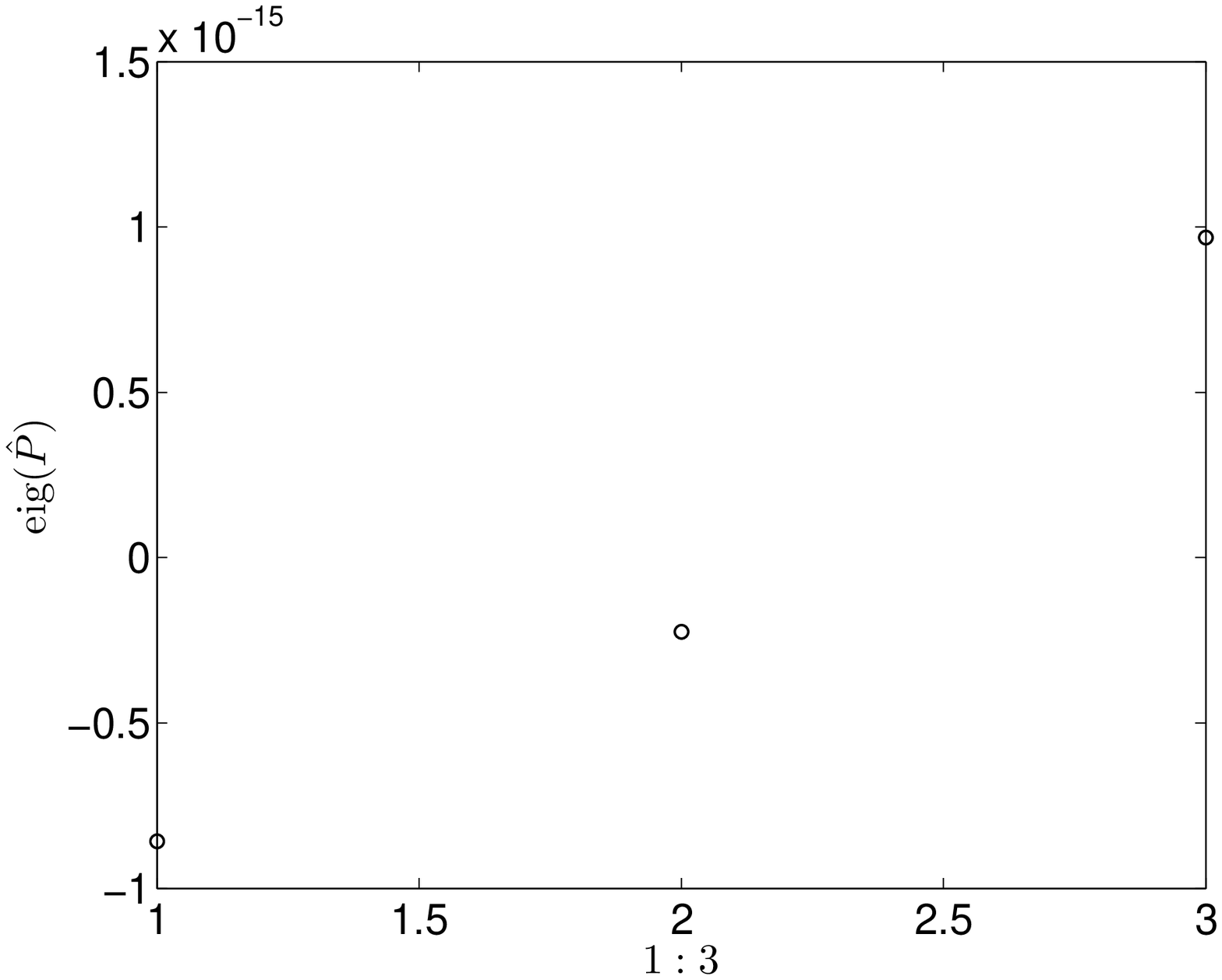}
\caption{Computed eigenvalues of operator $\hat{P}=I-\hat{M}^{T}\hat{M}^0$ (left) with $N_z = 56$ and $N_r = 24$ as presented in Example \ref{example_helium_radial}. $\hat{P}$ should be a projection operator thus we expect the eigenvalues to be either zero or one. We see, by investigating the region around zero, that operator $\hat{P}$ fulfills this property (right). \label{eigenvalues_hatP_radial}
}
\end{center}
\end{figure}

We start by checking the eigenvalues of $\hat{P}$ with this extended moment matrix. The eigenvalues should still be either zero or one since $\hat{P}$ is a projection operator. Figure \ref{eigenvalues_hatP_radial} shows the computed eigenvalues of this operator for Example \ref{example_helium_radial} which shows only eigenvalues equal to zero or one.

\subsection{Test Constrained Runs algorithm in Example \ref{example_helium_radial}}
This section tests the Constrained Runs algorithm on the model problem
presented in Example \ref{example_helium_radial}.  It is based on
parameters suggested by Gusarov et al.~\cite{gusarov} for the laser
ablation problem.  The boundary conditions of the Boltzmann
discretization are equilibrium distributions of the surface parameters
at the surface and equilibrium distributions of the ambient parameters
at the outer boundary. We perform 10000 Boltzmann time steps on the
initial state based on the ambient parameters presented in Example
\ref{example_helium_radial}. The distribution functions are rescaled
with the mass $m$ and weights $2\pi \Delta v_z \Delta v_r$,
corresponding to the radial velocity directions. The traveling
wave that emerges in the domain  moves from the surface, on the left, towards the
ambient boundary, at the right. The resulting reference distribution function is
denoted as $\fc$. The corresponding equilibrium distribution is
denoted as $\vect{f}^{eq}$.

We test the performance of the CR algorithm as a lifting operator that
maps density, average flow velocity, and temperature to distribution
functions. The CR algorithm is combined with Newton's method to ensure
stability. It uses a GMRES algorithm to invert the Jacobian matrix
in Newton's method and the matrix-vector product of the Jacobian
matrix is estimated using finite differences.

\begin{figure}[!htop]
\begin{center}
\begin{tabular}{cc}
& $m=0$\\
\includegraphics[width = 0.5\textwidth]{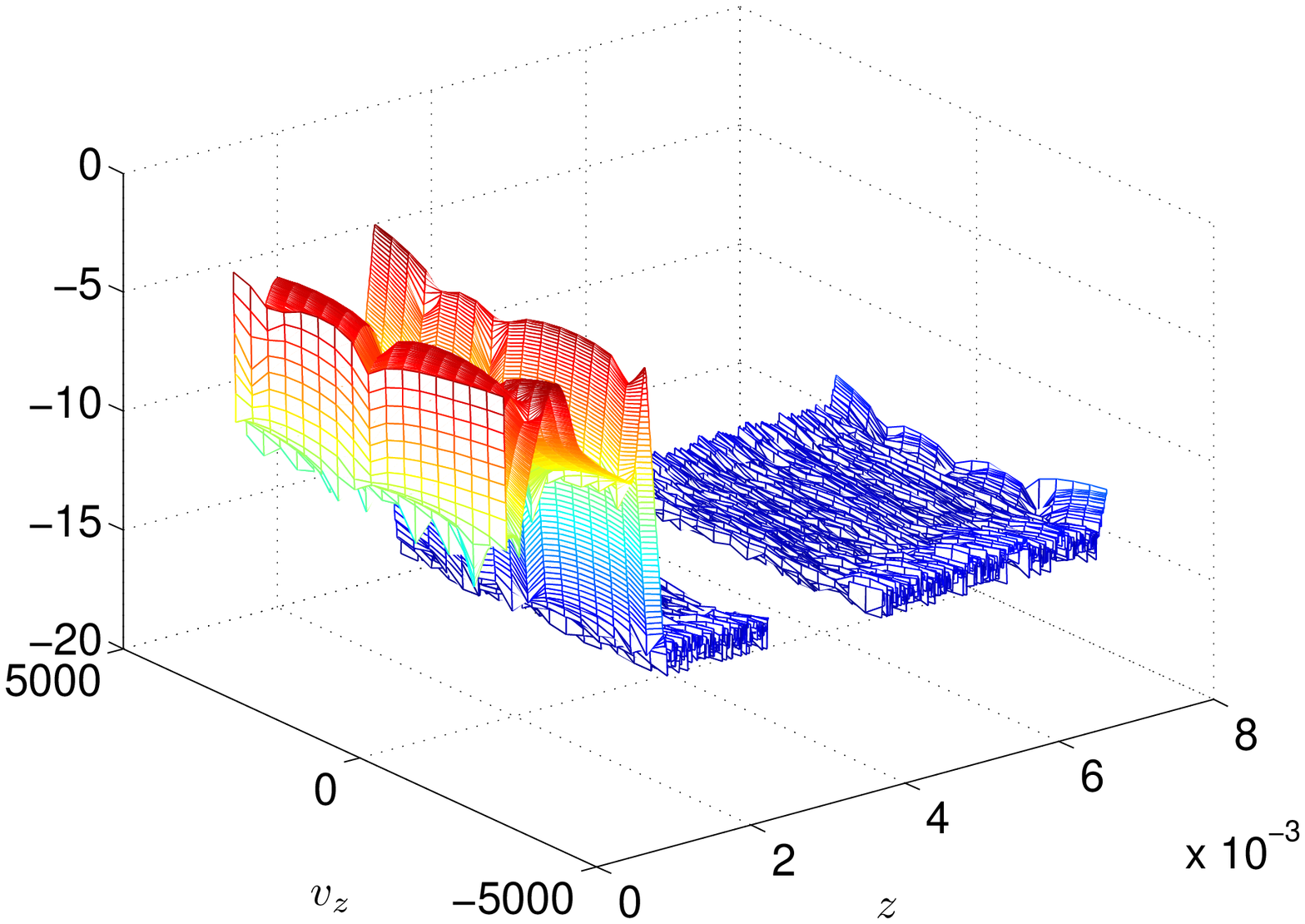} & \includegraphics[width=0.5\textwidth]{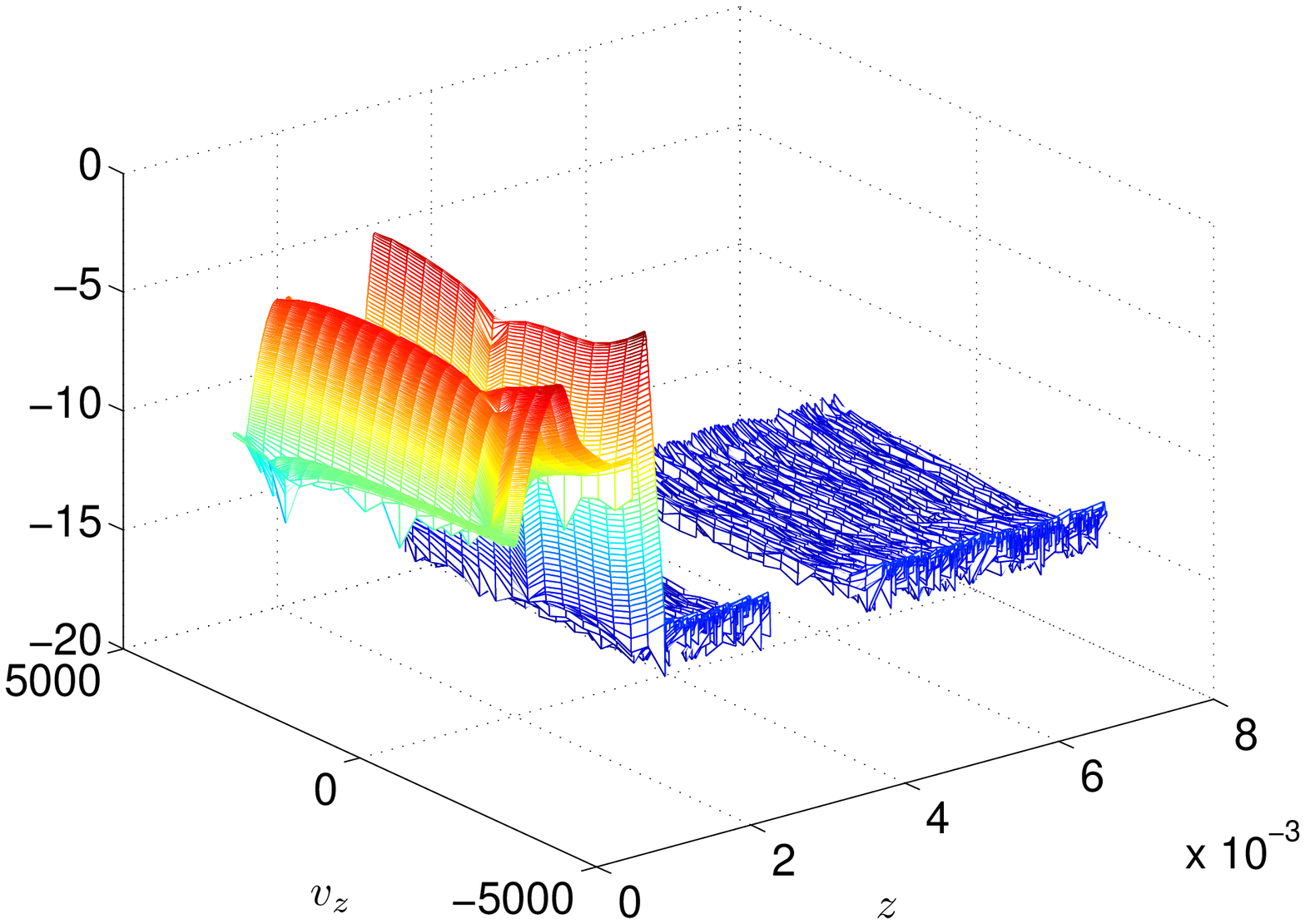} \\
$m=1$ & $m=2$ \\
\includegraphics[width = 0.5\textwidth]{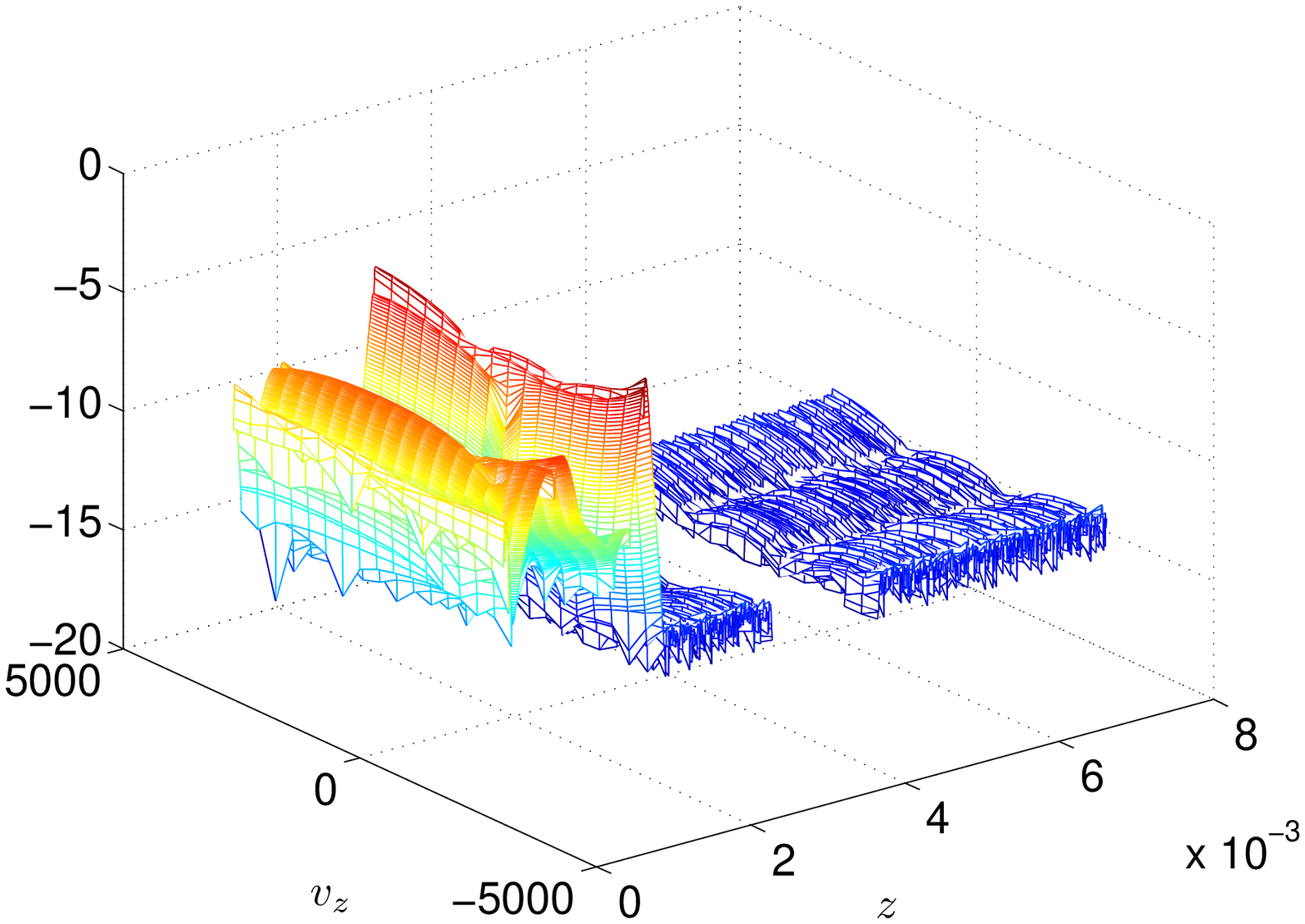} & \includegraphics[width=0.5\textwidth]{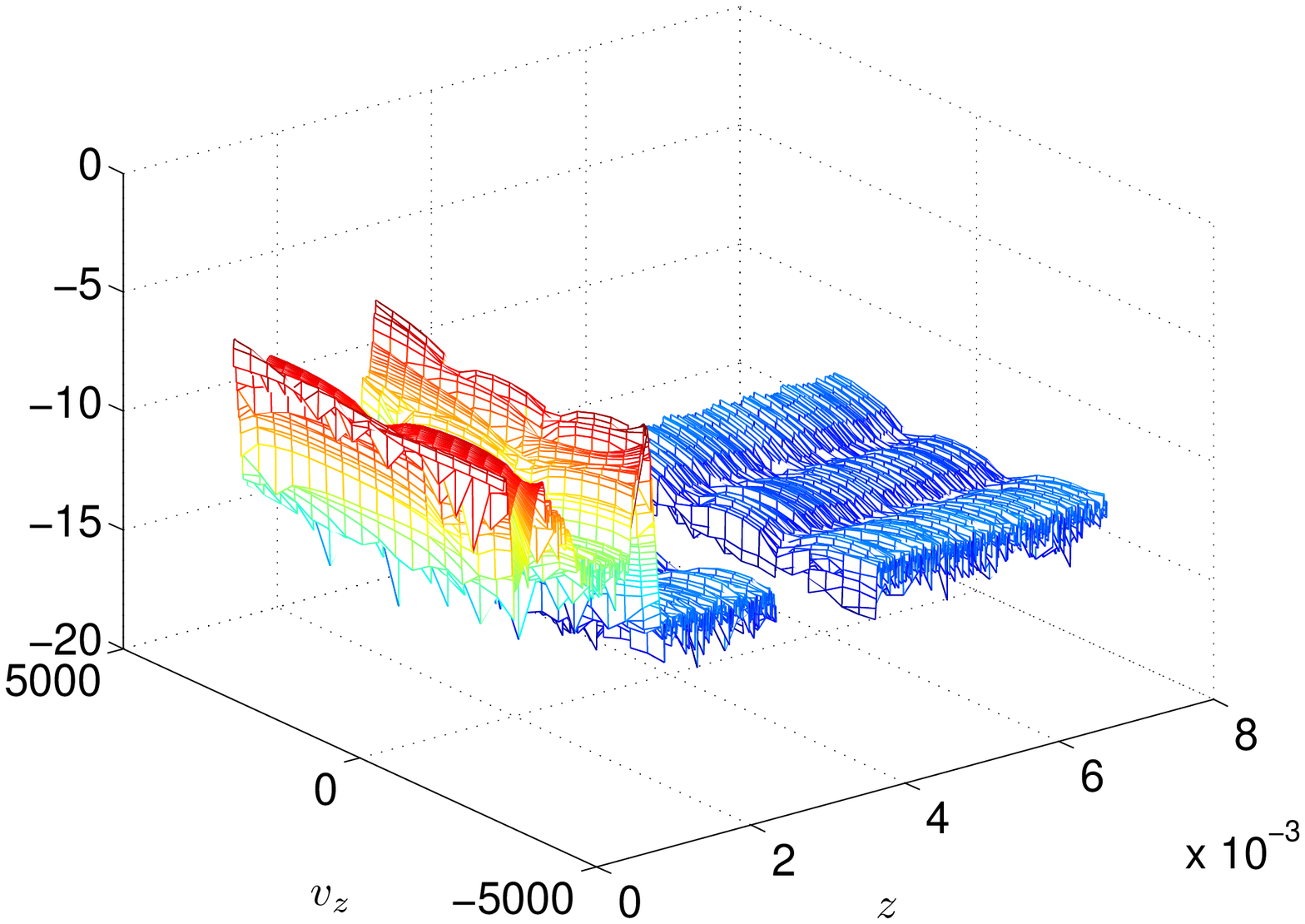} 
\end{tabular}
\caption{Log plot of the relative error $\left|\frac{\vect{f}^{eq}-\fc}{\fc} \right|$ (top left) and similarly of $\left|\frac{\vect{f}-\fc}{\fc} \right|$ where $\fc$ represents the reference distribution function after 10000 time steps on the initial state with parameters presented in Example \ref{example_helium_radial}. $\vect{f}^{eq}$ is the corresponding equilibrium distribution and $\vect{f}$ the distribution functions based on lifting with the Constrained Runs algorithm of order $m=0$ (top right), $m=1$ (bottom left), and $m=2$ (bottom right). Furthermore, the results include radial velocities $\vect{v}_r$. The errors are based on sums of distribution functions over the different radial velocity directions.
\label{fig_CR_helium2}
}
\end{center}
\end{figure}

Figure \ref{fig_CR_helium2} presents a log plot of the relative errors
$\left|\frac{\vect{f}^{eq}-\fc}{\fc} \right|$ (top left) and
$\left|\frac{\vect{f}-\fc}{\fc} \right|$. These results are plotted in
function of spatial grid points $\vect{z}$ and velocities in the
$z$-direction $\vect{v}_z$. $\vect{f}$ corresponds to the distribution
function based on lifting with the Constrained Runs algorithm of order
$m=0$ (top right), $m=1$ (bottom left), and $m=2$ (bottom right).  It
shows a narrow range of velocity directions $\vect{v}_z$ since the
outer velocities bring the distribution functions to
zero. Furthermore, the results include radial velocities
$\vect{v}_r$. The errors are based on sums of distribution functions
over the different radial velocity directions.

There are two remarks that illustrate the drawbacks of the Constrained Runs algorithm.


\begin{remark}
  A serious drawback is the dependence of the convergence rate of the
  Constrained Runs algorithm on the parameters of the Boltzmann
  model. This is illustrated in Figures \ref{projection} and
  \ref{projection2}. This makes it computationally not feasible to use
  the CR algorithm as a lifting operator for certain choices of the
  parameters of the Boltzmann model, especially when the grid
  resolution is smaller than the mean free path. This is illustrated in
  Figure \ref{conv_CR}, where the number of outer GMRES iterations is
  shown in function of $N$, the number of spatial grid points for
  $m=0$ (circle), $m=1$ (asterisk), $m=2$ (plus sign), $m=3$ (point),
  $m=4$ (cross), and $m=5$ (triangle). It is clear that the
  convergence rate of CR is determined by the parameters of the
  Boltzmann model, especially the ratio of $\Delta x$ to the
  mean free path $\lambda$, is important.
\begin{figure}[!htop]
\begin{center}
\includegraphics[width = 0.5\textwidth]{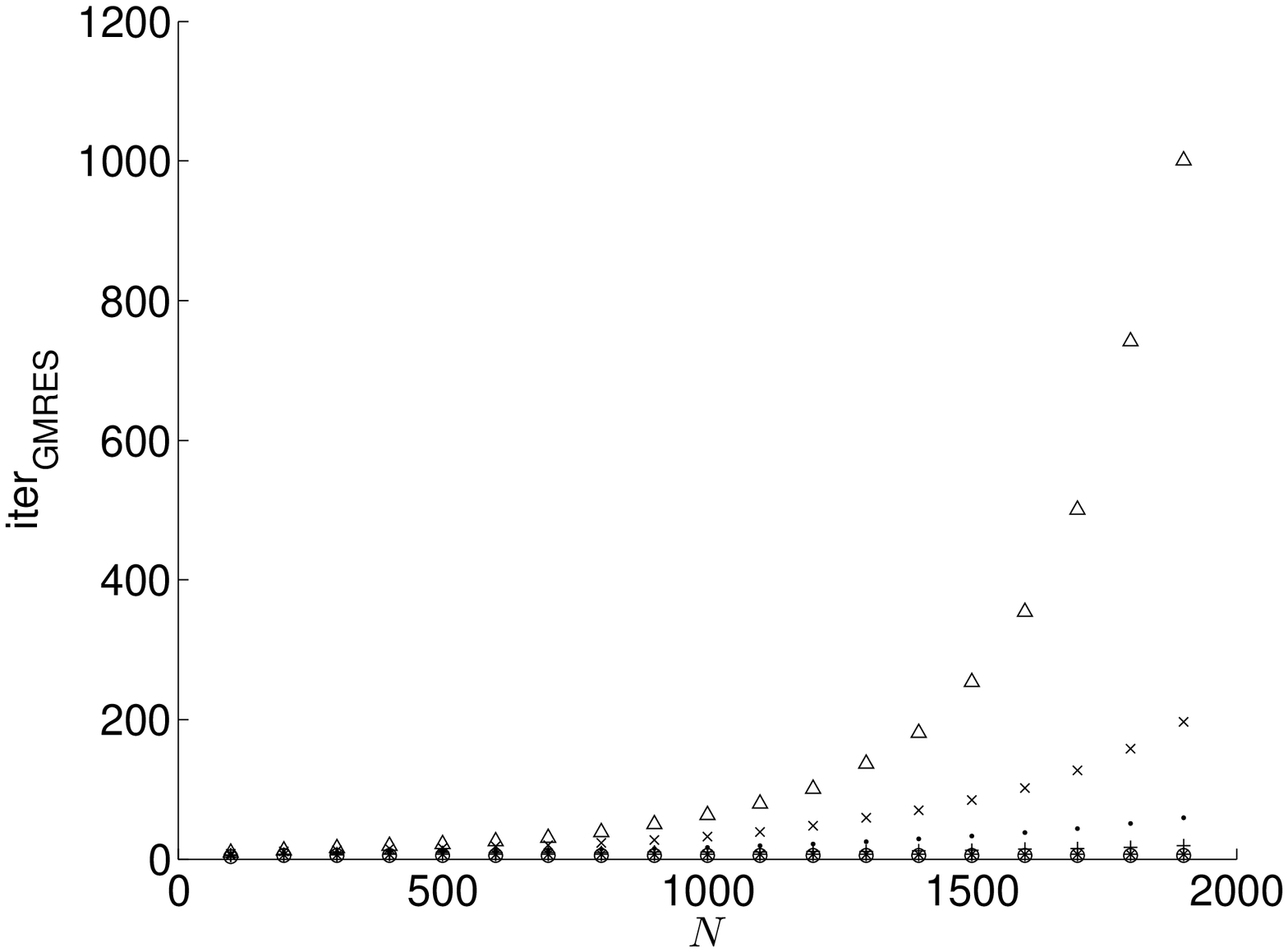}\includegraphics[width = 0.5\textwidth]{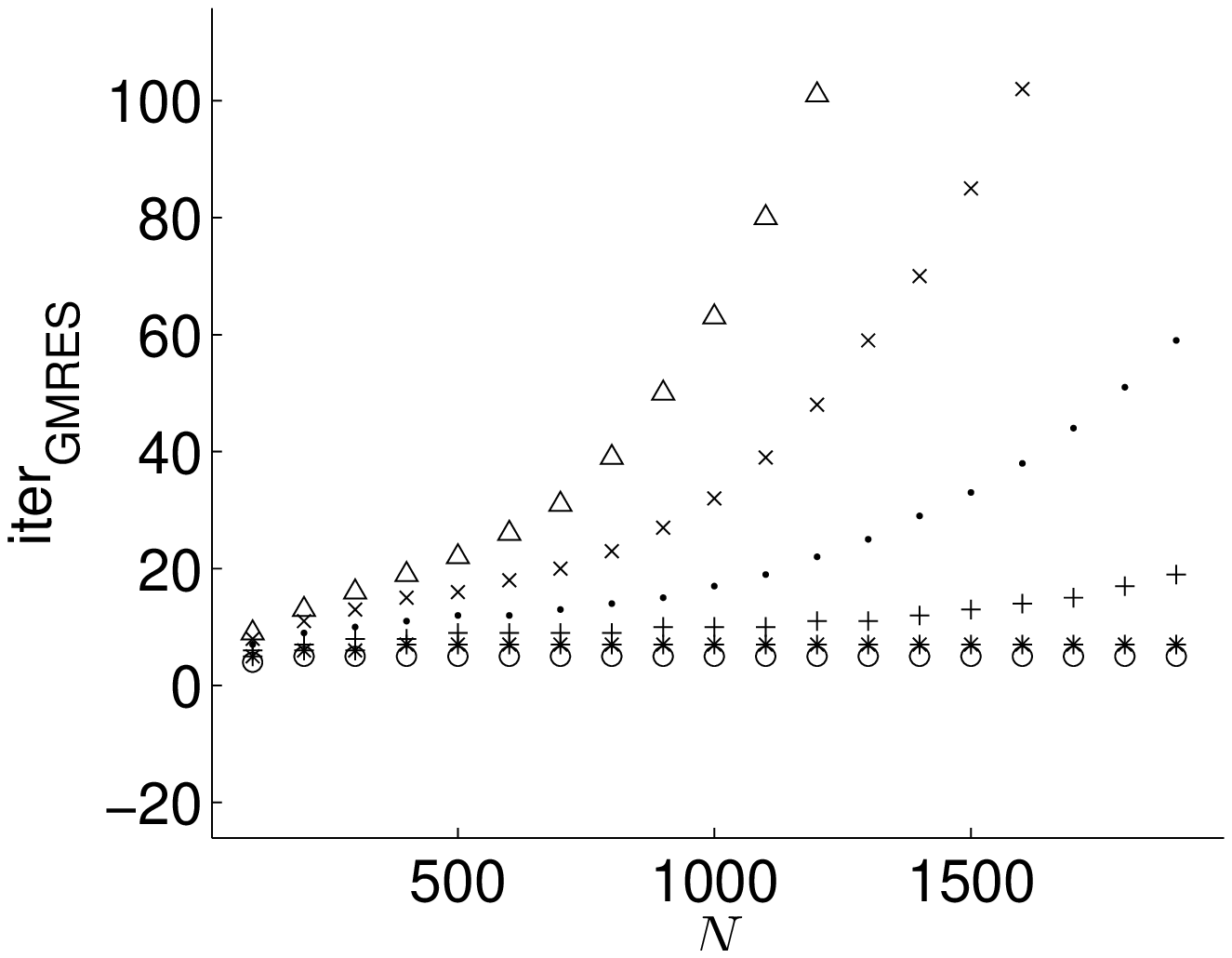}
\caption{Plot of the number of outer GMRES iterations in function of
  $N$, the number of spatial grid points for $m=0$ (circle), $m=1$
  (asterisk), $m=2$ (plus sign), $m=3$ (point), $m=4$ (cross), and
  $m=5$ (triangle). with parameters presented in Example
  \ref{example_helium}. The plot on the right represents a zoom of the
  left figure. This shows that the convergence rate of CR depends on
  the parameters of the Boltzmann model. \label{conv_CR}}
\end{center}
\end{figure}

To obtain a good convergence rate of Constrained Runs in the previous
examples, we chose the length of the domain $L = 30000\lambda$ in
\eqref{parameters_30000} and \eqref{parameters_30000_radial}. We would
rather like to have $L=30\lambda$ such that the nonequilibrium layer
is well described. It typically has a thickness of 10 to 20 mean free
path lengths \cite{gusarov}.  This means that the current grid step
equal to $H=30000\lambda/N$ has to be reduced to $h=30\lambda/N$. This
can be interpreted as refinements from Boltzmann$_H$ to Boltzmann$_h$.

However, to refine Boltzmann models additional operators are necessary
that transfer information between different scales.
There is some literature on the coupling of lattice Boltzmann models
with different grid resolutions. Guzik et al.~present a space-time
interpolation method to couple different grid resolutions for lattice
Boltzmann models \cite{guzik, guzik2}.  Here the first nonequilibrium
term is matched between the two Boltzmann models.

There is no literature for general discrete Boltzmann models with
multiple velocities, to the best of our knowledge.

If it is possible to transfer between the Boltzmann
subdomains with different grid sizes, denoted Boltzmann$_H$ and
Boltzmann$_h$, then the following scenario would be possible. 

First, there is a lifting operator that transfers information between
Boltzmann$_H$ and PDE$_H$ where $H$ is large. This has the
advantage that this is better for the convergence rate of the
Constrained Runs algorithm since the grid sizes determine the spectral
radius as highlighted in section \ref{sec_conv_CR}. Then the Boltzmann
model is systematically refined to reach the correct resolution to
resolve the non-equilibrium layer.  An overall analysis is required to
maintain the accuracy at each boundary.

\end{remark}


\begin{remark}
  Another drawback of the CR algorithm in general is the computational
  expense since it is based on performing Boltzmann steps to determine
  the Jacobian matrix in Newton's method.  In a similar way, when a
  matrix-free method is used such as GMRES the convergence is slow
  without preconditioning.
\end{remark}
\begin{remark}
  The Knudsen number $K_n$, which is the ratio of the mean free path
  and the feature length, is often used to classify
  flow regimes \cite{struchtrup}:
\begin{itemize}
\item{$K_n < 0.01$: hydrodynamic regime,}
\item{$0.01 < K_n < 0.1$: slip flow regime,}
\item{$0.1 < K_n < 10$: transition regime,}
\item{$K_n > 10$: free molecular flow.}
\end{itemize}
This classification is used as a rule of a thumb to determine
which mathematical description is necessary.

In laser ablation, the surface gets hot and ejects particles. At the
surface a nonequilibrium layer emerges that requires a full Boltzmann
description. While away from the surface, a PDE model can be used to
simulate the flow. Under what conditions can we lift information from the PDE model to
information required by the Boltzmann model?  Can this be linked to the
classification with the Knudsen number?

To know the link between the microscopic Boltzmann model and the
macroscopic PDE models, a Chapman--Enskog expansion is used. This
assumes that the Knudsen number is a small parameter (to achieve the
Navier--Stokes equations). Indeed, in the hydrodynamic regime, the
Knudsen number is a small parameter.

This Knudsen number should help us to determine the conditions when to
switch from Boltzmann to PDE model in the hybrid domain. However, this
is, so far, not investigated in detail since the Knudsen number
depends on the feature length that is hard to estimate from region
to region in the domain.
\end{remark}


\section{Conclusions} \label{conclusion}

Many problems based on the Boltzmann equation require the conversion of
moments to the corresponding distribution function. One motivating
example is laser ablation where a material plume is ejected from a
melting material that is heated by a laser.  To describe this process
accurately a lifting operator is required that maps hydrodynamic
moments, namely density, average flow velocity, and temperature, to
distribution functions of the kinetic Boltzmann model at the interface
between the melted material and the gas.

This paper extends the applicability of the Constrained Runs (CR)
algorithm that lifts these moments to distribution functions to
general discretizations of the Boltzmann equation.  Previously the CR
algorithm was only used to lift lattice Boltzmann models, where it
initializes or couples different models together. In this paper, we
have focused on lifting with CR in a finite volume discretization of
the Boltzmann equation, but we believe that the results of the 
CR algorithm can be applied to different discretizations.

The main difficulty encountered in this paper is that the
straightforward formulation of CR for Boltzmann models requires the
inverse of the moment matrix. This is a very ill-conditioned
matrix with the properties of a Vandermonde matrix and traditional numerical
methods to invert this matrix fail.  In this paper we have reformulated
the algorithm such that this inversion can be avoided all together.

With this new formulation the method can conserve multiple macroscopic
variables and include multiple velocity directions.  The paper
includes numerical results that test the restriction and lifting. We
perform 10000 Boltzmann time steps on the initial state of the laser
ablation problem to create a reference distribution function. This
reference solution is restricted to its macroscopic variables,
density, average flow velocity, and temperature, and lifted back to
distribution functions to test the generalized CR algorithm. The error
with the original reference distribution can now be reduced by
increasing the order of the CR algorithm.

A remaining drawback of the CR algorithm is the computational cost
since it is based on the time scale that a Boltzmann simulation needs
to reach the slow manifold. Another drawback is the dependence of the
convergence rate of CR on the parameters of the Boltzmann domain:
length of the domain expressed as mean free path lengths, step sizes,
$\ldots$

In \cite{vanderhoydonc2} and \cite{vanderhoydonc3} we have applied,
for LBMs, the CR on distribution functions represented as the first
few terms of the Chapman--Enskog expansion. The iteration then
determines the coefficients of the expansion rather than the
distribution function itself. This significantly reduces the size of
the problem and makes it easy to solve the implicit, higher order
problems. In the future, it might be possible to apply this technique
for general discretizations of Boltzmann models where multiple
moments are conserved since this paper generalized the underlying CR algorithm. However, when a general Maxwell--Boltzmann
equilibrium is used and multiple moments are given, the Chapman--Enskog
expansion becomes very complicated with the derivatives of the various
moments. At this moment it is unclear if this numerical Chapman--Enskog
technique is applicable in this context.

\section*{Acknowledgments}
This work is supported by research project \textit{Hybrid macroscopic
  and microscopic modelling of laser evaporation and expansion},
G.017008N, funded by `Fonds Wetenschappelijk Onderzoek' together with
an `ID-beurs' of the University of Antwerp. Furthermore, we would like to thank Annemie Bogaerts and David
Autrique of the Plasmant group of the University of Antwerp for their help in defining a model problem for the laser
ablation problem.

\end{document}